\documentclass[a4paper,11pt]{article}	
\pdfoutput=1
\usepackage{dcolumn}

\usepackage{bm}
\usepackage[T1]{fontenc}
\usepackage{graphicx}
\usepackage{amssymb,amsmath}
\usepackage{booktabs}
\usepackage{multirow}
\usepackage{cite,color,url}
\usepackage[dvipsnames]{xcolor}
\def\linkcolor{cyan!70!black}

\usepackage[
colorlinks=true
,urlcolor=\linkcolor
,anchorcolor=\linkcolor
,citecolor=\linkcolor
,filecolor=\linkcolor
,linkcolor=\linkcolor
,menucolor=\linkcolor
,linktocpage=true
,pdfproducer=medialab
,pdfa=true
]{hyperref}

\usepackage{slashed}
\usepackage{epsfig,psfrag,rotating,soul}
\usepackage{rotfloat}
\usepackage[font={small}]{caption}
\usepackage{colortbl}
\usepackage{soul}
\usepackage{xtab}
\usepackage{tikz}
\usepackage{makecell}

\usepackage{tabularx,threeparttable}

\evensidemargin \oddsidemargin
\usepackage{adjustbox}

\DeclareUnicodeCharacter{03BC}{\mu}

\def\be{\begin{equation}}
\def\ee{\end{equation}}
\def\ba{\begin{array}}
\def\ea{\end{array}}

\def\alambda{A_\lambda}
\def\akappa{A_\kappa}
\def\mueff{\mu_\mathrm{eff}}

\def\tanb{\tan\beta}

\def\sQ3{\widetilde{Q}_3}
\def\sU3{\widetilde{U}_3}
\def\sD3{\widetilde{D}_3}

\def\hsm{h_{\rm SM}}

\def\hs{h_{S}} 
\def\as{a_{S}}
\def\ntrli{\tilde\chi_{i}^0}

\def\ntrlone{\tilde\chi_{1}^0}
\def\ntrltwo{\tilde\chi_{2}^0}
\def\ntrlthree{\tilde\chi_{3}^0}
\def\ntrlfour{\tilde\chi_{4}^0}

\def\ntrltwothree{\tilde\chi_{{2,3}}^0}

\def\charonepm{\tilde\chi_{1}^\pm}
\def\charonemp{\tilde\chi_{1}^\mp}

\def\mone{M_1}
\def\mtwo{M_2}

\def\msinglino{m_{{\widetilde{S}}}}

\def\mntrlone{m_{{\tilde\chi}_{1}^0}}
\def\mntrltwo{m_{{\tilde\chi}_{2}^0}}
\def\mntrlthree{m_{{\tilde\chi}_{3}^0}}

\def\mcharone{m_{{\tilde\chi}_{1}^\pm}}

\def\mhsm{m_{h_{\mathrm{SM}}}}

\newcommand{\vev}[1] {\langle #1 \rangle}

\newcommand{\beq}{\begin{equation}} 
\newcommand{\eeq}{\end{equation}} 
\newcommand{\bea}{\begin{eqnarray}}  
\newcommand{\eea}{\end{eqnarray} }  
\newcommand{\bal}{\begin{align}}
\newcommand{\eal}{\end{align}}   
\newcommand{\bi}{\begin{itemize}}  
\newcommand{\ei}{\end{itemize}}  
\newcommand{\ben}{\begin{enumerate}}
\newcommand{\een}{\end{enumerate}}  
\newcommand{\bc}{\begin{center}}
\newcommand{\ec}{\end{center}} 
\newcommand{\bt}{\begin{table}}
\newcommand{\et}{\end{table}}  
\newcommand{\btb}{\begin{tabular}}
\newcommand{\etb}{\end{tabular}}

\def\vev{{\it vev}}

\def\vu{v_u}
\def\vd{v_d}

\def\vs{v_{\!_S}}
\def\pythia8{{\tt PYTHIA8}}

\def\z3nmssm{$Z_3$-NMSSM}
\allowdisplaybreaks

\let\OLDthebibliography\thebibliography
\renewcommand\thebibliography[1]{
  \OLDthebibliography{#1}
  \setlength{\parskip}{0pt}
  \setlength{\itemsep}{0pt plus 0.3ex}
}

\usepackage{fontawesome5}
\makeatletter
\newcommand{\github}[1]{%
   \href{#1}{\faGithubSquare}%
}
\makeatother

\usepackage{lipsum} 

\usepackage{tikz,xcolor,hyperref}

\definecolor{lime}{HTML}{A6CE39}
\DeclareRobustCommand{\orcidicon}{%
	\begin{tikzpicture}
	\draw[lime, fill=lime] (0,0) 
	circle [radius=0.16] 
	node[white] {{\fontfamily{qag}\selectfont \tiny ID}};
	\draw[white, fill=white] (-0.0625,0.095) 
	circle [radius=0.007];
	\end{tikzpicture}
	\hspace{-3mm}
}

\foreach \x in {A, ..., Z}{%
	\expandafter\xdef\csname orcid\x\endcsname{\noexpand\href{https://orcid.org/\csname orcidauthor\x\endcsname}{\noexpand\orcidicon}}
}

\begin{document}

\begin{titlepage}

\thispagestyle{empty}

\def\thefootnote{\fnsymbol{footnote}}

\begin{flushright}
EFI-25-12\\
IFT-UAM/CSIC-25-93\\
\end{flushright}

\vspace*{0.5cm}

{\Large\flushleft\sffamily\bfseries\par
Shedding Light on Dark Matter at the LHC with Machine Learning
}

\vspace{0.25cm}
\hrule height 1.5pt
\vspace{0.25cm}

{\bfseries\raggedright\sffamily\par
Ernesto~Arganda\orcidA{}$^{1, 2}$%
\footnote{{\tt \href{mailto:ernesto.arganda@uam.es}{ernesto.arganda@uam.es}}}%
, Mart\'in~de~los~Rios\orcidB{}$^{3,4}$%
\footnote{\tt \href{mailto:mdelosrios@unc.edu.ar}{mdelosrios@unc.edu.ar}}%
, Andres D. Perez\orcidC{}$^{2}$%
\footnote{\tt \href{mailto:andresd.perez@csic.es}{andresd.perez@csic.es}}%
, Subhojit~Roy\orcidD{}$^{5}$%
\footnote{\tt \href{mailto:sroy@anl.gov}{sroy@anl.gov}}%
, Rosa~M.~Sand\'a Seoane\orcidE{}$^{1, 2}$%
\footnote{{\tt \href{mailto:rosa.sanda@uam.es}{rosa.sanda@uam.es}}}%
\, and Carlos~E.~M.~Wagner\orcidF{}$^{5, 6, 7, 8, 9}$%
\footnote{{\tt \href{mailto:cwagner@uchicago.edu}{cwagner@uchicago.edu}}}%
}

\vspace{0.1cm}

{\sl\footnotesize
\begin{flushleft}
$^1$Departamento de Física Teórica, Universidad Autónoma de Madrid, Cantoblanco, 28049 Madrid, Spain

$^2$Instituto de Física Teórica UAM-CSIC, C/ Nicol\'as Cabrera 13-15, Universidad Autónoma de Madrid,\\ \, Campus de Cantoblanco, 28049 Madrid, Spain

$^3$SISSA -  International School for Advanced Studies, Via Bonomea 265, 34136 Trieste, Italy

$^4$Instituto de Astronom\'ia Te\'orica y Experimental, CONICET - UNC, Laprida 854, X5000BGR, C\'ordoba, Argentina

$^5$HEP Division, Argonne National Laboratory, 9700 Cass Ave., Argonne, IL 60439, USA

$^6$Enrico Fermi Institute, Physics Department, University of Chicago, Chicago, IL 60637, USA

$^7$Kavli Institute for Cosmological Physics, University of Chicago, Chicago, IL 60637, USA

$^8$Leinweber Center for Theoretical Physics, University of Chicago, Chicago, IL 60637, USA 

$^9$Perimeter Institute for Theoretical Physics, Waterloo, Ontario N2L 2Y5, Canada
\end{flushleft}
}

\vspace{0.06cm}

\renewcommand*{\thefootnote}{\arabic{footnote}}
\setcounter{footnote}{0}

\noindent 
{\sc Abstract:} We investigate a WIMP dark matter (DM) candidate in the form of a singlino-dominated lightest supersymmetric particle (LSP) within the $Z_3$-symmetric Next-to-Minimal Supersymmetric Standard Model (NMSSM). This framework gives rise to regions of parameter space  where DM is obtained via co-annihilation with nearby higgsino-like electroweakinos and DM direct detection~signals are  suppressed, the so-called ``blind spots''. On the other hand, collider signatures remain promising due to enhanced radiative decay modes of higgsinos into the singlino-dominated LSP and photons, rather than into leptons or hadrons. Compared to MSSM scenarios with light bino- and wino-like electroweakinos, the NMSSM allows for final states with multiple photons arising from cascade radiative decays, providing a distinctive collider signature. This motivates searches for radiatively decaying neutralinos, however, these signals face substantial background challenges, as the decay products are typically soft due to the 
small mass-splits ($\Delta m$) between the LSP and the higgsino-like coannihilation partners. We apply a data-driven Machine Learning (ML) analysis that improves sensitivity to these subtle signals, offering a powerful complement to traditional search strategies to discover a new physics scenario. 
Using an LHC integrated luminosity of $100~\mathrm{fb}^{-1}$ at $14~\mathrm{TeV}$, the method achieves a $5\sigma$ discovery reach for higgsino masses up to $225~\mathrm{GeV}$ with $\Delta m\!\lesssim\!12~\mathrm{GeV}$, and a $2\sigma$ exclusion up to $285~\mathrm{GeV}$ with $\Delta m\!\lesssim\!20~\mathrm{GeV}$. These results highlight~the power of collider searches to probe DM candidates that remain hidden from current~direct detection experiments, and provide a motivation for a search by the LHC collaborations~using~ML~methods.

\end{titlepage}

\tableofcontents

\section{Introduction}
\label{sec:Intro}

The discovery of the Higgs boson ($\hsm$) at a mass near 125 GeV~\cite{ATLAS:2012yve, CMS:2012qbp}  marked a major milestone in particle physics, completing the particle spectrum of the Standard Model (SM) and confirming the mechanism of electroweak symmetry breaking. However, despite this success, the SM leaves several key questions unresolved, the nature of   Dark Matter (DM) and the stability of the electroweak scale, which is sensitive to radiative corrections induced by heavy particles interacting with the Higgs field.
Among the various extensions of the SM, low-scale supersymmetry (SUSY)~\cite{Fayet:1976et,Fayet:1977yc,Nilles:1983ge,Haber:1984rc,Gunion:1984yn,Martin:1997ns} remains one of the most compelling candidates. It offers a unified framework that simultaneously accommodates a viable DM particle~\cite{Farrar:1978xj,Dimopoulos:1981zb,Weinberg:1981wj,Sakai:1981pk,Dimopoulos:1981dw}, protects the electroweak scale from destabilizing quantum corrections, and provides solutions to other theoretical challenges of the SM, including the electroweak vacuum’s potential instability and the imperfect unification of gauge couplings.
In this framework, the lightest supersymmetric particle (LSP) becomes a viable DM candidate, provided that a discrete symmetry known as R-parity is conserved.

Although the LHC has not yet observed any clear signals of physics beyond the Standard Model (BSM) as it enters its high-luminosity phase, one of its central goals remains the continued search for new phenomena, along with high-precision measurements of the Higgs boson’s properties. An especially interesting direction is to revisit regions of parameter space where small statistical excesses have already been observed, using new data from the current/upcoming runs.
To improve search sensitivity in various challenging scenarios, such as scenarios involving compressed spectra or soft final states, the use of machine learning (ML) techniques has become increasingly important (for reviews of ML in high-energy physics see, for instance, Refs.~\cite{Larkoski:2017jix,Mehta:2018dln,Guest:2018yhq,Albertsson:2018maf,Radovic:2018dip,Carleo:2019ptp,Bourilkov:2019yoi,Feickert:2021ajf,Schwartz:2021ftp,Karagiorgi:2021ngt,Coadou:2022nsh,Shanahan:2022ifi,Plehn:2022ftl,Belis:2023mqs,Bardhan:2024zla,Mondal:2024nsa,Halverson:2024hax}). These methods enable the exploitation of complex correlations between observables that are often inaccessible to traditional cut-based analyses. Here, we adopt an ML-based strategy to enhance the LHC discovery potential in some BSM scenarios where conventional approaches typically suffer from low signal efficiency due to the softness of visible decay products.

In this study, we focus on a well-motivated DM scenario within the $Z_3$-symmetric Next-to-Minimal Supersymmetric Standard Model (NMSSM)~\cite{Ellwanger:2009dp}, where the LSP is a singlino-dominated neutralino. Owing to its stability, neutrality, and weak-scale interactions, this state naturally qualifies as a Weakly Interacting Massive Particle (WIMP) type DM~\cite{Steigman:1984ac,Arcadi:2017kky}. However, a pure singlino tends to underperform in early-universe annihilation processes, leading to an overabundance of relic DM. 
To ensure compatibility with the observed relic density from Planck data~\cite{Planck:2018vyg}, the singlino-dominated LSP must efficiently annihilate or co-annihilate in the early universe. These processes can proceed via $s$-channel resonance through $Z$ or $\hsm$, $t$-channel exchanges of charginos/neutralinos, or co-annihilations with close-in-mass electroweakinos-particularly higgsinos and, to a lesser extent, binos and winos~\cite{Arkani-Hamed:2006wnf}. In our analysis, we work in a regime where sleptons are heavy and decoupled, suppressing lepton superpartner-assisted co-annihilation.

Meanwhile, DM direct detection (DMDD) experiments such as XENONnT~\cite{XENON:2023cxc}, PandaX-4T~\cite{PandaX-4T:2021bab}, and LUX-ZEPLIN (LZ)~\cite{LZ:2022lsv, LZCollaboration:2024lux} have significantly improved bounds on spin-independent (SI) and spin-dependent (SD) DM-nucleus scattering, placing stringent constraints on the couplings between the LSP and Higgs or gauge bosons. In the singlino-dominated DM case, these couplings can be strongly suppressed, leading to so-called \emph{blind spots} in DMDD~\cite{Badziak:2015exr, Baum:2017enm}.
Traditionally, DMDD-SI scattering cross-section blind spots have been linked to scenarios where the singlino-higgsino interaction term with $\hsm$, governed by $\lambda$, is pretty small, often requiring a scenario with $\kappa > 0$. Ref.~\cite{Roy:2024yoh} recently demonstrated that new blind spots can also occur for $\kappa < 0$, due to a cancellation between the bino-higgsino-$\hsm$ and the singlino-higgsino-$\hsm$ contributions when the relative signs of $\mueff$, $M_1$, and the singlino mass term $2\kappa\mueff/\lambda$ are appropriately chosen. This opens up new regions of parameter space that remain allowed by current DMDD limits, or where the predicted rates fall below the so-called neutrino floor, leaving these regions beyond the reach of even future DMDD experiments.

In this work, we investigate a compressed electroweakino spectrum in the NMSSM, where the next-to-lightest supersymmetric particle (NLSP) is a higgsino-like neutralino exhibiting sizable radiative decay into the singlino-dominated LSP and a photon, i.e., $\ntrltwo \rightarrow \ntrlone \gamma$. We identify a previously underexplored region in which both higgsino-like states, $\ntrltwo$ and $\ntrlthree$, possess enhanced branching ratios (BR) for sequential radiative decays: $\ntrlthree \rightarrow \ntrltwo \gamma$ followed by $\ntrltwo \rightarrow \ntrlone \gamma$. These decay chains can yield striking photon-rich signatures at the LHC.
Importantly, such decay modes have not been the focus of conventional LHC electroweakino searches, which primarily target wino-like states decaying via on-shell $Z$ or $\hsm$ into final states such as $3\ell + E_T^\text{miss}, 1\ell + 2b + E_T^\text{miss}$, or $2\ell + E_T^\text{miss}$~\cite{ATLAS:2020ckz,CMS:2017moi,CMS:2018szt,ATLAS:2018ojr}, where $E_T^\text{miss}$ is the missing transverse energy (MET). While additional searches have been designed to probe compressed spectra through soft leptons or initial state radiation (ISR) jets~\cite{ATLAS:2021moa,CMS:2021few,ATLAS:2022zwa}, many decay modes relevant to our scenario remain untested, allowing for significantly relaxed mass bounds. As a result, current bounds on higgsino-like states may be significantly weaker in this scenario.  Moreover, several SUSY searches featuring photons exist, their signal regions are not optimized to search for soft photons in the final state~\cite{CMS:2018fon,CMS:2019vzo,ATLAS:2022ckd,CMS:2023xlp}, and thus do not directly constrain our parameter space of interest. The singlino-higgsino parameter space considered here resides within spin-independent DM scattering blind spots, involving a bino-like state nearby in mass as the fourth neutralino that plays a negligible role in collider observables due to its suppressed production cross section. These features underscore the importance of complementary collider probes of such elusive DM scenarios.

\begin{figure}
\centering
\includegraphics[width=0.49\textwidth]{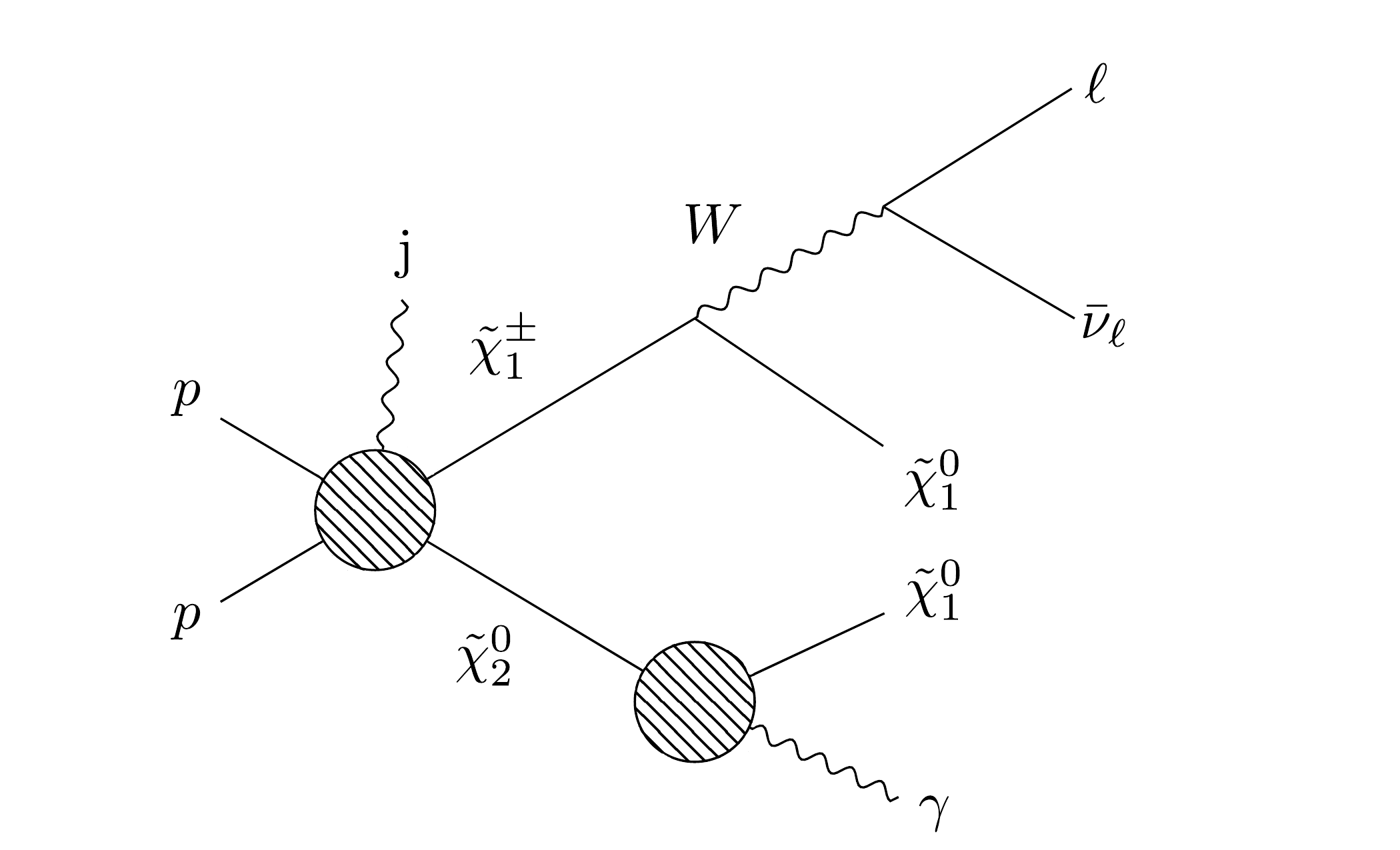}
\includegraphics[width=0.38\textwidth]{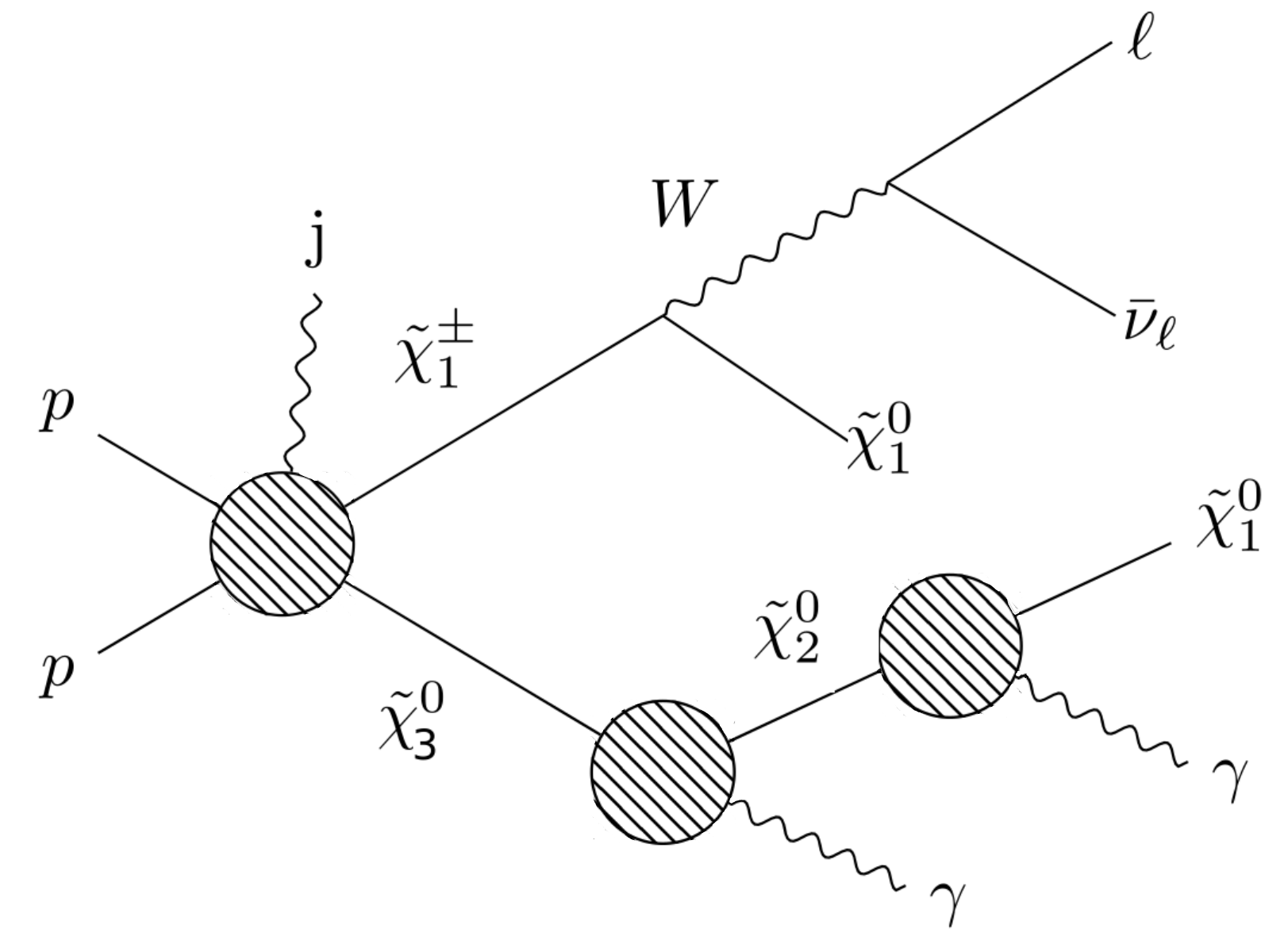}
\caption{Feynman diagrams of the two production channel modes of the higgsino-like states, $p p \rightarrow \ntrltwo \charonepm j$ (left) and $p p \rightarrow \ntrlthree \charonepm j$ (right).}
\label{fig:feynman-diagram}
\end{figure}

 Let us emphasize that although the region of parameter we are exploring is somewhat fine tuned, this fine tuning is dictated not by theoretical arguments but by the current strong experimental constraints that demand the spin independent direct detection cross section to be suppressed in regions of parameters like the compressed region, which leads to the proper Dark Matter relic density. This scenario is therefore well motivated and deserves a detailed study of its experimental signatures.
Regarding the corresponding search strategy, we employ ML techniques to probe radiative neutralino decays at the LHC.  ML techniques have already been employed to explore other regions of the NMSSM parameter space, concentrating on the explanation of several anomalies appearing in LHC data~\cite{Hammad:2025wst}.~\footnote{An analysis explaining the same anomalies in our framework would be interesting but would require a separate dedicated study.}

We focus on proton--proton collisions at a center-of-mass energy of $\sqrt{s} = 14$~TeV and a total integrated luminosity of $\mathcal{L} = 100\,\text{fb}^{-1}$. This setup reflects an early projection for Run 3 of the LHC.  We note that the actual Run 3 energy is 13.6 TeV; however, the small difference in
center-of-mass energy has only a minor impact on the relevant kinematic distributions and does
not affect the qualitative conclusions of our analysis nor the comparative performance of the ML
methods. The modest luminosity choice is motivated by the need to reduce computational demands associated with large-scale Monte Carlo simulations. For the ML-based analysis, we assess the discovery potential using two approaches: the standard Binned-Likelihood (BL) method~\cite{Cowan:2010js} and the more recent Machine-Learned Likelihoods (MLL) framework~\cite{Arganda:2022qzy,Arganda:2022mrd,Arganda:2022zbs,Arganda:2023qni}. In the MLL approach, the likelihood function is modeled using Kernel Density Estimators (KDE)~\cite{RosenblattKDE,ParzenKDE} applied to the output of the ML classifier, enabling an unbinned fit that captures fine details of the probability distribution.
Since the compressed mass scenario leads to soft visible final states, we consider the production of higgsino-like $\ntrltwothree \charonepm$ pairs in association with a hard ISR jet~\cite{Roy:2024yoh}, as shown in the Feynman diagrams in Figure~\ref{fig:feynman-diagram}. The ISR jet boosts the system and enhances the $E_T^\text{miss}$ signature, which is crucial for suppressing the SM background. While simple cut-and-count analyses struggle in this compressed regime, we find that ML techniques offer significantly improved sensitivity and could prove essential for electroweakino discovery prospects at the LHC in the near future.

 Let's stress that the study of the photon signatures induced by radiatively decaying neutralinos in the NMSSM have never been performed before. Second, compared to a previous analysis in the MSSM~\cite{Arganda:2024tqo}, the signatures are different, related to different decay chains in the two scenarios. For instance, as we will show, there is a novel signature associated with the possibility of having multiple photons in the final state. The number of photons, indeed, becomes one of the most significant variables in determining the significance of detection in the Machine Learning (ML) analysis. Furthermore, the compressed-mass
regime leads to distinctive collider signatures that are particularly challenging and well suited for ML techniques. Although the ML tools employed are similar to those used in the MSSM analysis performed in Ref.~\cite{Arganda:2024tqo}, each signal comes with its own characteristics and challenges, and in this case the
new final state and multiplicities merit a dedicated and detailed study, since its implementation and more importantly its feasibility and reach is not trivial to infer a priori.

This article is organized as follows: in Section~\ref{thrysetup} we introduce the theoretical framework considered, the NMSSM. We highlight the relevant parameter space in the neutralino sector, and discuss the phenomenology of the singlino-higgsino compressed mass scenario. In Section~\ref{sec:collider} we describe the particular signal and background, with the corresponding event simulation that we employ for our collider analysis. Section~\ref{sec:ML} is devoted to the characterization of the machine learning procedures, both the BL and the unbinned MLL methods used for the final significance estimation. The main results of our analysis are shown in Section~\ref{sec:results}. Finally, Section~\ref{sec:conclusions} summarizes our conclusions. 

\section{Theoretical Framework and Dark Matter Phenomenology}
\label{thrysetup}

\subsection{The \texorpdfstring{$Z_3$}{Z3}-Symmetric NMSSM Setup}

As mentioned in the introduction, several works have been conducted in the context of singlino-dominated DM in the $Z_3$-symmetric NMSSM~\cite{Ellwanger:2009dp, Cao:2016nix,Ellwanger:2016sur,Xiang:2016ndq,Baum:2017enm,Cao:2018rix,Ellwanger:2018zxt,Domingo:2018ykx,Baum:2019uzg,vanBeekveld:2019tqp, Abdallah:2019znp,Cao:2019qng,Abdallah:2020yag,Lopez-Fogliani:2021qpq,Kim:2021suj,Kim:2022gpl,Datta:2022bvg, Das:2012rr,Ellwanger:2014hia,Chatterjee:2022pxf,Cao:2022htd,Cao:2023juc,Roy:2024yoh,Bisal:2023mgz, Guchait:2020wqn, Cao:2022ovk, Adhikary:2024dix, Meng:2024lmi, Badziak:2015exr, Cheung:2014lqa, Badziak:2017uto, Ellwanger:2004xm, Fan:2011yu, Ellwanger:1996gw, Belanger:2005kh, Gunion:2005rw, Cheung:2013dua, Fan:2012jf, Berlin:2014pya, Draper:2010ew, Cerdeno:2004xw, Cao:2015loa,  Cerdeno:2007sn, Cao:2021tuh, Hugonie:2007vd, Bi:2015qva, Cheung:2012pg, Franke:2001nx, Kim:2014noa, Potter:2015wsa, Barman:2020vzm, Wang:2009rj, Adhikary:2022pni, Zhou:2021pit}, which extends the Minimal Supersymmetric Standard Model (MSSM) by including a gauge singlet superfield $\widehat{S}$~\cite{Ellwanger:2009dp}.
The superpotential of the $Z_3$-symmetric NMSSM is given by~\cite{Ellwanger:2009dp}
\beq
{\cal W}= {\cal W}_\mathrm{MSSM}|_{\mu=0} + \lambda \widehat{S}
\widehat{H}_u \cdot \widehat{H}_d
        + \frac{\kappa}{3} \widehat{S}^3 \, ,
\label{eqn:superpot}
\eeq
where ${\cal W}_\mathrm{MSSM}|_{\mu=0}$ is the MSSM superpotential including the Yukawa interactions of the Higgs doublet superfields with the SM quark and leptons superfields, but with no
higgsino mass term (known as $\mu$-term), $\widehat{H}_u, \widehat{H}_d$ are the $SU(2)$ Higgs doublet superfields of the MSSM, respectively, and `$\lambda$' and `$\kappa$' are dimensionless coupling constants.  Adding the extra superfield $\widehat{S}$ to the MSSM solves the so-called $\mu$-problem~\cite{Kim:1983dt} when it gets a non-zero vacuum expectation value~(\vev)~$\vs$ and generates dynamically an effective $\mu$-term given by $\mueff=\lambda \vs$ from the second term in Eq.~\eqref{eqn:superpot}. The soft SUSY-breaking Lagrangian is given by
\beq
-\mathcal{L}^{\rm soft}= -\mathcal{L_{\rm MSSM}^{\rm soft}}|_{B\mu=0}+ m_{S}^2
|S|^2 + (
\lambda A_{\lambda} S H_u\cdot H_d
+ \frac{\kappa}{3}  A_{\kappa} S^3 + {\rm h.c.}) \,,
\label{eqn:lagrangian}
\eeq
where $m_S$ denotes the soft SUSY-breaking mass associated with the singlet scalar field `$S$', and $\alambda$, $\akappa$ are the dimensionful trilinear soft-breaking parameters characteristic of the NMSSM. The term $\mathcal{L_{\rm MSSM}^{\rm soft}}|_{B\mu=0}$ includes no $H_u H_d$ Higgs bilinear soft supersymmetry breaking term.

The electroweakino sector contains the neutralinos and the charginos.
The NMSSM neutralino sector is enriched by a fifth state, the singlino, arising from the fermionic component of the gauge singlet superfield $\widehat{S}$.
The symmetric $(5\times 5)$ neutralino mass matrix, in the basis $\psi^0=\{\widetilde{B},~\widetilde{W}^0, ~\widetilde{H}_d^0,
~\widetilde{H}_u^0, ~\widetilde{S}\}$, is given by~\cite{Ellwanger:2009dp}
\beq
\label{eqn:mneut}
{\cal M}_0 =
\left( \begin{array}{ccccc}
\mone & 0 & -\dfrac{g_1 \vd}{\sqrt{2}} & \dfrac{g_1 \vu}{\sqrt{2}} & 0 \\[0.4cm]
\ldots & \mtwo & \dfrac{g_2 \vd}{\sqrt{2}} & -\dfrac{g_2 \vu}{\sqrt{2}} & 0 \\
\ldots & \ldots & 0 & -\mueff & -\lambda \vu \\
\ldots & \ldots & \ldots & 0 & -\lambda \vd \\
\ldots & \ldots & \ldots & \ldots & 2 \kappa \vs
\end{array} \right) \,,
\eeq
where 
$g_1$ and $g_2$ are the $U(1)_Y$ and 
$SU(2)_L$ gauge couplings, respectively, and $\vu=v\sin\beta$, $\vd=v\cos\beta$ such that
$v^2=\vu^2+\vd^2 \approx (174$~GeV)$^2$ and $\tan\beta=\vu/ \vd$.
$\mone$ ($\mtwo$) is the soft SUSY-breaking masses for the $U(1)_Y$ ($SU(2)_L$) gaugino, known as bino (wino).
The (5,5) element of the neutralino mass-matrix in Eq.~\eqref{eqn:mneut} is the singlino mass term, $\msinglino = 2\kappa \vs$. 

The symmetric matrix ${\cal M}_0$ in the absence of $CP$-violation can be diagonalized by an orthogonal $(5 \times 5)$ matrix `$N$',
i.e.,
\bea
\label{eqn:diagonalise-1}
N {\cal M}_0 N^T = {\cal M}_D = {\rm diag}(m_{\tilde\chi_1^0},m_{\tilde\chi_2^0},m_{\tilde\chi_3^0},m_{\tilde\chi_4^0},m_{\tilde\chi_5^0})  \, . 
\eea
The neutralino mass-eigenstates ($\ntrli$) are represented in terms of the weak eigenstates by ($\psi_j^0$)
\beq
\ntrli = N_{ij} \psi_j^0 \,,
\label{eqn:diagN2}
\eeq
where $i$ and $j$ run from 1 to 5.  At tree level, the electroweakino sector is defined by six fundamental input parameters:  $\{ \, \tanb, \, \lambda, \, \kappa \, \,M_1, \, M_2, \, \mueff\}$.
 In this work, we focus on the scenario where singlino-dominated neutralino is the lightest supersymmetric particle (LSP). The singlino state mixes with the other neutralino states like bino and higgsinos and leads to the well-known well-tempered singlino-dominated neutralino LSP~\cite{Feng:2000gh, Ellwanger:2009dp}. This admixture plays a crucial role in determining the DM relic abundance, direct detection rates, and collider phenomenology.
\subsection{Dark Matter Relic Density and Co-Annihilation}

In this work, we focus on scenarios where the lightest neutralino $\ntrlone$ is singlino-dominated, while the next-to-lightest states $\ntrltwo$ and $\charonepm$ are higgsino-like. 
To comply with the Planck upper bound on the DM relic density ($\Omega_{DM} h^2 \lesssim 0.12$)~\cite{Planck:2018vyg}, the singlino-dominated LSP must annihilate efficiently in the early universe. The NMSSM provides several mechanisms to achieve this, including $s$-channel annihilation via the $Z$ boson, the SM-like Higgs ($\hsm$), or light singlet-like scalar and pseudoscalar states ($h_S$, $a_S$); $t$-channel exchange of charginos or heavier neutralinos; and co-annihilation with nearby electroweakinos.

Co-annihilation processes dominate when the mass gap is small enough to prevent thermal decoupling of the NLSP,
which contributes to the annihilation processes before freeze-out.  These contributions are Boltzmann-suppressed by $\exp[-(m_{X_2} - \mntrlone)/T]$ but can be offset by larger annihilation cross-sections of the co-annihilating species.
Additional enhancement can arise from ``assisted co-annihilation''~\cite{Ellwanger:2016sur}, where $X_2$ (the co-annihilating particle) annihilates via a resonant Higgs-mediated process, especially when $m_{X_2} \approx m_H/2$.
The efficiency of these processes depends on the mixing of the singlino with other neutralinos, controlled by parameters such as $\lambda$, $\kappa$, $M_1$, and $\mu_{\text{eff}}$, considering the decoupled wino scenario of the electroweakino sector, that is, large $M_2$.  In particular, $Z$- and $\hsm$-mediated channels require a non-negligible higgsino admixture, while co-annihilation requires small mass splittings between the LSP and the NLSP.
Here, we focus on the ``compressed region'' of parameter space, where the DM relic density is primarily achieved through the co-annihilation mechanism, as the mass difference between the singlino-dominated LSP and the higgsino-like states is assumed to be relatively small.
The degree of compression can be parameterized via:
\begin{equation}
\varepsilon \equiv \frac{\mntrltwo}{\mntrlone} - 1 \,. 
\end{equation}

\subsection{Direct Detection  Blind Spots and Collider Phenomenology}

As emphasized before,
direct detection  experiments have achieved remarkable sensitivity, leading to increasingly stringent upper limits on the spin-independent (SI) and spin-dependent (SD) DM-nucleus elastic scattering cross-sections~\cite{XENON:2019rxp, XENON:2023cxc, PandaX-4T:2021bab, LZ:2022lsv, PICO:2017tgi, IceCube:2021xzo, Fermi-LAT:2017bpc, MAGIC:2016xys, LZCollaboration:2024lux}. These bounds impose notable constraints on the interactions between the DM and  the SM particles, particularly through couplings to the SM-like Higgs boson and $Z$ boson, which are 
primarily responsible for the DMDD SI and SD scattering cross-sections, respectively.
The singlino-dominated DM around the electroweak scale, satisfying these constraints, often requires the theory to be close to so-called ``blind spot'' configurations, where the relevant couplings are suppressed.
It is well known that, in the limit of minimal mixing between the SM-like Higgs and other Higgs states, and considering only the singlino-higgsino $(3\times3)$ neutralino neutralino sector, a blind spot in the SI cross-section appears when the mass ratio $\mntrlone/\mueff > 0$ which corresponds to $\kappa > 0$~\cite{Badziak:2015exr, Baum:2017enm, Cheung:2014lqa, Badziak:2017uto, Ellwanger:2004xm}. In such a regime, the singlino-higgsino-Higgs coupling, which is proportional to $\lambda$, becomes effectively suppressed, reducing the direct detection cross-section.
Recently, a comprehensive set of blind spot conditions for both SI and SD DM-nucleon scattering in the context of singlino-dominated DM has been derived, focusing on the $(4\times4)$ neutralino mass matrix involving the bino, higgsinos, and singlino states, assuming the wino is decoupled~\cite{Roy:2024yoh, Abdallah:2020yag}.
Ref.~\cite{Roy:2024yoh} identifies a novel blind spot condition for singlino-like DM that emerges through mixing effects with bino and higgsino states.
The SI scattering is dominated by Higgs exchange, and a blind spot occurs when the coupling to the SM-like Higgs vanishes,
$g_{h_{\rm SM} \, \ntrlone \ntrlone} \approx 0$, which corresponds to the 
 the blind spot condition:
\bea
\label{eqn:coupling-blindspot-approxi}
\bigg(\mntrlone + \frac{g_1^2 v^2}{M_1 - \mntrlone}\bigg) \frac{1}{\mueff \sin 2\beta} \simeq 1 \,.
\eea
This condition can be satisfied even for $\kappa < 0$, expanding the viable parameter space beyond the standard singlino-higgsino blind spot scenario.
This feature appears when the singlino mass term $ 2\kappa \mueff/\lambda$ and $\mueff$ have opposite signs. For $\lambda > 0$, this corresponds to the region where $\kappa < 0$. The underlying mechanism involves a cancellation between two competing contributions to the lightest neutralino: the gaugino-higgsino-Higgs interaction, driven by $g_1$, and the singlino-higgsino-Higgs coupling, controlled by $\lambda$.
Consequently, the blind spot condition is generally governed by the relative sign between $\mueff$ and $M_1$, requiring them to be opposite for $\kappa > 0$, and aligned for $\kappa < 0$. This interplay gives rise to a new viable region in the parameter space where the spin-independent direct detection cross-section of singlino-dominated DM is significantly reduced.
The sign combinations among $\kappa$, $\mueff$ and $M_1$ that are preferred for satisfying the blind spot condition, shown in Eq.~\eqref{eqn:coupling-blindspot-approxi}, are listed in Table~\ref{tab:signcomb}~\cite{Roy:2024yoh}.
\begin{table}[t]
\renewcommand{\arraystretch}{1.6}
\begin{center}
\begin{tabular}{|c|c|c|c|c|}
\hline
$\kappa$ & $\mueff$ & $M_1$ \\
\hline
\multirow{2}{*}{$-$} & $+$  & $+$ \\
\cline{2-3}
 &  $-$  & $-$ \\
\hline
\multirow{2}{*}{$+$} &$+$  & $-$ \\
\cline{2-3}
 &   $-$  & $+$ \\
\hline
\end{tabular}
\caption{Blind spot condition satisfying, pointed out in Eq.~\eqref{eqn:coupling-blindspot-approxi}, preferred sign combinations among $\kappa$, $\mueff$, and $M_1$. }
\label{tab:signcomb}
\end{center}
\end{table} 

Novel collider signatures can be observed in these blind spot regions.
The singlino-like LSP can co-annihilate with nearby higgsino-like states, particularly in a compressed mass spectrum where $\ntrltwo$ and $\ntrlthree$ are close in mass to $\ntrlone$. In this regime, conventional three-body decays such as $\ntrltwo \to \ntrlone f \bar{f}$ are suppressed by $\varepsilon^5$, while loop-induced radiative decays like $\ntrltwo \to \ntrlone \gamma$ are less suppressed, scaling as $\varepsilon^3$~\cite{Ambrosanio:1996gz}. As a result, the radiative modes dominate and offer distinctive collider handles.
The size of these radiative branching ratios is controlled by the electroweakino mixing structure. In particular, suppressed $g_{\tilde\chi_i^0 \ntrlone Z}$ and $g_{\tilde\chi_i^0 \ntrlone h}$ couplings—common in the low-$\lambda$ blind spot regime—enhance the dipole transitions to photons. Furthermore, the presence of a light bino-like state ($\ntrlfour$) close in mass to the singlino and higgsino states can enhance radiative transitions such as $\ntrltwo \to \ntrlone \gamma$ and $\ntrlthree \to \ntrltwo \gamma$, while simultaneously reducing the DMDD scattering cross-section. Thus, bino admixture plays a crucial dual role in collider and DM phenomenology.

From a collider perspective, one of the most promising signatures of this compressed spectrum is the production channel $pp \to \ntrltwothree \charonepm$, followed by radiative decays of $\ntrltwo$ or $\ntrlthree$ and a leptonic $W$ decay from $\charonepm$. This leads to $1\ell + \gamma + E_T^\text{miss}$ final states, with typically soft leptons and photons due to the small mass gaps. While ATLAS has conducted diphoton searches~\cite{ATLAS:2020qlk}, they assume on-shell Higgs decays ($h \to \gamma\gamma$) and are not sensitive to the off-shell radiative transitions relevant to our setup.
In cases where the visible decay products are too soft to trigger, mono-jet or mono-photon + $E_T^\text{miss}$ channels become important, enabled by ISR-tagged production of higgsino-like states~\cite{ATLAS:2018nud, CMS:2021far}. While these searches do provide constraints in the compressed region, they remain largely insensitive to singlino pair production due to its small cross-section.

This compressed blind spot region is particularly attractive because it remains consistent with current and projected direct detection constraints. In some cases, the DMDD-SI cross-section falls below the neutrino floor, making collider searches the only feasible discovery probe. The presence of soft leptons and photons with large missing transverse energy, often alongside a hard ISR jet, motivates advanced analysis strategies. In this work, we explore the use of machine learning (ML) techniques to improve signal discrimination and enhance sensitivity to this challenging but phenomenologically rich parameter space. Our collider analysis strategy is detailed in Section~\ref{sec:collider}.
%

\subsection{Constraints and the Studied Parameter Space}
\label{constraints}
We impose both theoretical and experimental constraints to ensure the viability of the parameter space. Theoretical requirements include the absence of tachyons, vacuum stability, and perturbativity of couplings up to high scales, etc. These are evaluated using \texttt{NMSSMTools}~\cite{Ellwanger:2005dv,Das:2011dg}, which also computes the particle spectrum and relevant low-energy observables.
Dark-matter observables are computed with \texttt{MicrOMEGAs}~\cite{Belanger:2006is,Belanger:2008sj,Barducci:2016pcb} interfaced to \texttt{NMSSMTools}.
We do not impose the Planck value $\Omega_{\rm DM}h^2=0.120$~\cite{Planck:2018vyg} as an equality constraint when selecting the parameter space.
Instead, we treat it as an upper bound and allow late-time entropy injection, such as from a strong first-order phase transition~\cite{Chatterjee:2022pxf, Ghosh:2022fzp, Bittar:2025lcr, Hooper:2025fda, Roy:2025zvo} or  from the decay of a long-lived field, to dilute~\cite{Roy:2022gop} otherwise overabundant benchmarks to the observed value, consistent with BBN provided the reheating temperature satisfies $T_{\rm RH}\gtrsim\text{few MeV}$. 
This assumption does not affect collider signatures.

Direct detection bounds on spin-independent (SI) and spin-dependent (SD) scattering cross sections are applied using results from LZ~\cite{LZCollaboration:2024lux}, XENONnT~\cite{XENON:2019rxp,XENON:2023cxc}, PandaX-4T~\cite{PandaX-4T:2021bab}, PICO-60~\cite{PICO:2017tgi}, and IceCube~\cite{IceCube:2021xzo}.
Finally, indirect detection constraints on DM annihilation to the SM final states are included from Fermi-LAT~\cite{Fermi-LAT:2017bpc} and MAGIC~\cite{MAGIC:2016xys}, limiting the thermal cross sections into $\mu$, $\tau$, $b$, and $\hsm$, $\gamma$, $W$ final states.
Experimental bounds from Higgs physics are incorporated via \texttt{HiggsBounds}~\cite{Bechtle:2020pkv} and \texttt{HiggsSignals} \cite{Bechtle:2020uwn}, ensuring consistency with LEP, Tevatron, and LHC searches, as well as the signal strengths of the 125 GeV Higgs boson. We allow its mass to lie within $122$--$128$ GeV, accounting for theoretical uncertainties.

Models with light electroweakinos may be constrained by their contributions to the anomalous magnetic moment of the muon ($a_{\mu} = (g_{\mu}-2)/2$).
In our setup, the contributions to $a_{\mu}$ can be rendered small by considering heavy sleptons. Due to 
recent theoretical determinations of the SM prediction of  $a_{\mu}$, which suggest a small deviation with respect to the SM prediction~\cite{Muong-2:2006rrc, Muong-2:2021ojo, Muong-2:2025xyk, Aliberti:2025beg}, we consider heavy sleptons in this work. Larger contributions, however, could be obtained by lowering the mass of the sleptons, which do not play a relevant role in our analysis.

Searches for electroweakinos at the LHC primarily target multilepton and missing energy final states resulting from $\charonepm \ntrltwo$ and $\charonepm \charonemp$ production. These signatures are most efficiently probed when $\charonepm$ and $\ntrltwo$ are wino-like and nearly degenerate, while the LSP is bino-like. In such cases, the dominant decay modes are $\ntrltwo \to Z \ntrlone$ and $\charonepm \to W^\pm \ntrlone$, producing clean final states like $3\ell + E_T^\text{miss}$ or $2\ell + E_T^\text{miss}$~\cite{CMS:2018szt, ATLAS:2020ckz, CMS:2021edw}. Additional modes like $1\ell + 2b + E_T^\text{miss}$ are also considered when $\ntrltwo$ decays via an on- or off-shell Higgs~\cite{ATLAS:2018qmw, CMS:2019pov, ATLAS:2020pgy}.
However, these constraints weaken significantly in scenarios where the electroweakinos are higgsino-like. In such cases, both the production cross section and the branching ratios into $Z$ and $h$ bosons are reduced. This is especially true in regions where the singlino or bino admixture alters decay patterns, or when alternative channels like $\ntrltwo \to \ntrlone \gamma$ dominate. 
The bounds further weaken in the compressed regime, where the mass difference between the NLSP and the LSP is below $m_Z$. Here, cascade decays produce off-shell electroweak bosons, leading to final states with low-$p_T$ leptons, jets, and missing energy~\cite{ATLAS:2019lng, CMS:2020bfa, ATLAS:2021moa, CMS:2021cox}. As $\Delta m$ decreases, detection efficiency drops, and the electroweakino mass limits degrade accordingly. The reduced leptonic branching ratios and softer kinematics challenge standard LHC search strategies~\cite{Roy:2024yoh}.

In the context of the $Z_3$-symmetric NMSSM, this situation becomes more subtle. Singlino-dominated LSPs and the presence of light singlet-like scalars can significantly alter decay chains, as shown in prior studies~\cite{Abdallah:2019znp, Abdallah:2020yag, Chatterjee:2022pxf, Datta:2022bvg}. These models permit novel topologies that are not well constrained by existing analyses. Notably, the compressed regions of parameter space are also favored in parts of the DMDD blind spot parameter space.

As mentioned before, we focus on such compressed scenarios where higgsino-like neutralinos undergo radiative decays like $\ntrltwo \to \ntrlone \gamma$ and $\ntrlthree \to \ntrltwo \gamma$, resulting in final states that include soft photons and missing energy. These signatures lie beyond the scope of many current LHC searches, offering a viable and interesting region for further exploration.
To assess the LHC viability of our benchmark points (BPs), we utilize both \texttt{SModelS}~\cite{Alguero:2021dig} and \texttt{CheckMATE}~\cite{Dercks:2016npn}, which recast existing LHC analyses including recent results with 139 fb$^{-1}$ of data. Both packages compute an exclusion ratio
$r = \frac{S - 1.64 \, \Delta S}{S_{95}}$
where $S$ is the predicted signal yield, $\Delta S$ is the associated Monte Carlo uncertainty, and $S_{95}$ is the 95\% CL experimental limit. Points with $r > 1$ are excluded. To account for higher-order corrections, we apply a conservative $k$-factor of 1.25 to all electroweakino production cross sections~\cite{Fiaschi:2018hgm}.

The ATLAS and CMS collaborations have independently carried out multiple analyses aimed at probing electroweakinos in the compressed mass spectrum regime~\cite{ATLAS:2021moa, ATLAS:2019lng, CMS:2021cox, CMS:2021edw}. Interestingly, both experiments have reported mild excesses, up to about $2.5\sigma$, in searches targeting chargino-neutralino production in compressed scenarios, particularly in channels involving soft di-leptons. These excesses may be interpreted as hints of higgsino-like states with mass splittings between the chargino and the lightest neutralino (LSP) in the range of approximately 5--20 GeV.
Such compressed mass spectra naturally arise in scenarios involving singlino-higgsino co-annihilation, as discussed in this work. In addition to the di-lepton signatures, our study highlights the possibility of alternative discovery channels involving photons. It would therefore be compelling for the LHC collaborations to explore this scenario using dedicated analyses with existing Run-2 data, particularly in photon-enriched final states, to investigate whether similar excesses are present. Regardless, the compressed electroweakino regime is expected to be a focus of further scrutiny in upcoming LHC runs, and it remains to be seen whether these mild excesses persist with more data.
At the same time, the current LHC analyses impose meaningful constraints on such compressed spectra. For example, the ATLAS study in Ref.~\cite{ATLAS:2019lng} and the CMS analysis in Ref.~\cite{CMS:2021cox} are already implemented in the \texttt{CheckMATE} framework~\cite{Lara:2025cpm}. These analyses target electroweakino production with final states containing two, three, or four leptons plus $E_T^\text{miss}$. A key strength of these searches lies in their inclusive strategy: they define multiple signal regions categorized by lepton multiplicity, flavor, and charge combinations, thereby maximizing sensitivity across a wide range of models.
We utilize \texttt{CheckMATE} to recast the constraints from these analyses onto the parameter space explored in this work. The resulting bounds are shown in Figure~\ref{fig:contour_plots} in Section~\ref{sec:collider}. As discussed earlier, we observe a notable weakening of the limits in our model due to the suppression of effective branching ratios into leptonic final states, a direct consequence of the singlino-higgsino admixture and the resulting decay patterns. 
For this reason, the present analysis should be viewed as complementary to
the previous LHC compressed spectrum analyses, focusing on the multi-lepton signature~\cite{ATLAS:2021moa, ATLAS:2019lng, CMS:2021cox, CMS:2021edw}.
This relaxation opens up parameter space that would otherwise be excluded in pure higgsino or wino scenarios.

To explore the parameter space, we fix several Lagrangian parameters to the values listed in Table~\ref{scanranges}. We vary $\kappa \, (< 2 \, \lambda)$  accordingly to scan the regions relevant for this work, with emphasis on small mass gaps between the singlino-dominated LSP and the higgsino-like states. We also vary $\mueff$ in order to explore the relevant higgsino mass range. 
The so-called soft trilinear coupling in the top sector, $A_\text{top}$, and the third-generation soft masses, $M_{Q_3}$ and $M_{U_3}$, are chosen such that the BPs reproduce the observed Higgs mass, $\mhsm = 125$~GeV. Within the selected parameter space, the remaining NMSSM trilinear soft terms, $\alambda$ and $\akappa$, are set so that the masses of the doublet-like heavy Higgs bosons ($H, A, H^{\pm}$) are larger than 1 TeV to comply with the collider constraints and the singlet-like $CP$-even and $CP$-odd Higgs bosons ($\hs$ and $\as$) are below 300~GeV. These Higgs bosons do not affect the collider analysis presented here; however, they do influence the DMDD-SI cross-section of the singlino-dominated LSP. 
To comply with the latest constraints and remain near the blind-spot condition defined in Eq.~\eqref{eqn:coupling-blindspot-approxi}, we take $M_1 \simeq 500$~GeV. All benchmark scenarios used in our study are listed in listed in Tables~\ref{table:bps2} and~\ref{table:bps3} in Appendix~\ref{sec:BPs}. 
\begin{table}[t]
\begin{center}
\begin{tabular}{|c|c|c|c|c|c|c|c|c|c|}
\hline
$\lambda$ & $\kappa$ & $\tanb$& \makecell{$- \mueff$ \\ (GeV)}&  \makecell{$-\alambda$ \\ (TeV)} &
\makecell{$\akappa$ \\ (GeV)}
 & \makecell{$\mone$ \\ (GeV)} & \makecell{$A_\text{top}$ \\ (TeV)} & \makecell{$M_{Q_{3}}$ \\ (TeV)} & \makecell{$M_{U_{3}}$ \\ (TeV)} \\
\hline
0.027 &0.010--0.0133&
6.2& $ 130-320 $ & 0.8-1.1 & 100 & 500 & 5 & 2.2 & 4.2\\
\hline
\end{tabular}
\caption{Fixed Lagrangian parameters used to select benchmark points, listed in Tables~\ref{table:bps2} and~\ref{table:bps3}. The parameter $\mueff$ is varied between $130$ and $320$~GeV to cover the higgsino-dominated neutralino–chargino mass range. The coupling $\kappa < 2 \lambda$ is varied accordingly to adjust the singlino-dominated LSP and the higgsino-dominated states for a given $\mueff$.}
\label{scanranges}
\end{center}
\end{table} 

\section{Collider Analysis}
\label{sec:collider}
%
As described in the previous section, the proposed signal has contributions from two different processes, depicted in Figure~\ref{fig:feynman-diagram}. The first channel involves the production of the lightest chargino, $\tilde\chi_1^{\pm}$, and the second-lightest neutralino, $\tilde\chi_2^0$, (both being higgsino-like), in addition to a highly energetic ISR jet, which recoils against the pair system. In the second channel, $\tilde\chi_1^{\pm}$ is produced in conjunction with the third-lightest neutralino, $\tilde\chi_3^0$ (also higgsino-like), that decays subsequently in the second-lightest neutralino $\tilde\chi_2^0$ and a soft photon, $\tilde\chi_3^0 \to \tilde\chi_2^0 + \gamma$. In both processes, the chargino \( \tilde\chi_1^{\pm} \) decays into the lightest neutralino $\tilde\chi_1^0$ (which is predominantly singlino-like) along with a lepton and a neutrino via an off-shell \( W \) boson, \( \tilde\chi_1^{\pm} \to \tilde\chi_1^0 \ell \nu_\ell \). The off-shell nature of the \( W \) boson arises due to the compressed mass spectrum, which prevents the on-shell production of the \( W \) mediator. Simultaneously, the second-lightest neutralino undergoes a radiative decay into the lightest neutralino and a photon, \( \tilde\chi_2^0 \to \tilde\chi_1^0 + \gamma \). Accordingly, the studied LHC signatures are described by:

\begin{equation}
\begin{array}{c}
pp \to \tilde\chi_1^{\pm} \, \tilde\chi_3^0 \, j \to \tilde\chi_1^0 \, \ell \, \nu_{\ell} + \tilde\chi_2^0 \, \gamma + j \to \tilde\chi_1^0 \, \ell \, \nu_{\ell} + \tilde\chi_1^0 \, \gamma \gamma + j \, , \\[8pt]
pp \to \tilde\chi_1^{\pm} \, \tilde\chi_2^0 \, j \to \tilde\chi_1^0 \, \ell \, \nu_{\ell} + \tilde\chi_1^0 \, \gamma + j \, .
\end{array}
\end{equation}

For event simulation, samples were produced using {\sc MadGraph5\_aMC@NLO}~\cite{Alwall:2014hca}, at leading order (LO) in QCD. The parton distribution functions were taken from the NNPDF2.3 LO set~\cite{Ball:2012cx}, and simulations were performed for proton-proton collisions at a center-of-mass energy of 14~TeV. Parton shower evolution and hadronization were carried out with {\sc Pythia8}~\cite{Sjostrand:2014zea,Sjostrand:2007gs}, while detector effects were approximated using the {\sc Delphes} fast simulation package~\cite{deFavereau:2013fsa}, employing the default ATLAS detector configuration~\footnote{Results are expected to be equally valid for CMS, as no detector-specific features significantly affect the signal topology or analysis strategy.}.  Since our signal requires a hard ISR jet, all relevant background samples were generated including additional jets in the matrix element.~Matrix-element~partons~were matched to the {\sc Pythia8} parton shower using the MLM matching scheme as implemented in {\sc MadGraph5\_aMC@NLO}, with $xqcut = 30$ GeV, in order to avoid~double~counting~of~hard~radiation.

\begin{table}[ht]
\centering

\begin{tabular}{c|c}
\textbf{Object} & \textbf{ID Criteria} \\
\hline
Electron & $p_T > 10$~GeV,\quad $|\eta|<2.47$, excluding $1.37<|\eta|<1.52$ \\
Muon & $p_T > 10$~GeV,\quad $|\eta|<2.7$ \\
Jet & $p_T > 20$~GeV,\quad $|\eta|<4.5$ \\
$\tau_h$ & $p_T > 20$~GeV,\quad $|\eta|<2.47$, excluding $1.37<|\eta|<1.52$ \\
Photon & $p_T > 10$~GeV,\quad $|\eta|<2.37$ \\
\end{tabular}
\caption{Object identification criteria used in the analysis.}
\label{table:objectid}
\end{table}

For signal events, we used for generation the UFO~\cite{Darme:2023jdn} implementation of the NMSSM model~\cite{Christensen:2008py,Duhr:2011se}. Spin correlations in the chargino decays $\tilde\chi_1^\pm \to \tilde\chi_1^0 \, \ell^\pm \,\nu_{\ell}$ ($\ell=e, \mu$) were accounted for via the decay chain functionality in {\sc MadGraph5\_aMC@NLO}. The simulated mass points span the range $m_{\tilde\chi_2^{0}}\in [150,290]$~GeV, with $m_{\tilde\chi_2^0} - m_{\tilde\chi_1^0} \in [3, 30]$ GeV 
(all benchmark points are summarized in Tables~\ref{table:bps2} and~\ref{table:bps3} of the Appendix~\ref{sec:BPs}).

The primary background corresponds to mono-photon final states characterized by significant missing transverse momentum, a high-$p_T$ ISR jet, and the presence of at least one charged lepton. The dominant contributions arise from $W+$jets, $W\gamma$, and $t\bar{t}+$jets. We also include subdominant processes such as $Z+$jets, single-top, $t\bar{t}\gamma$, and diboson production ($WW$, $ZZ$, and $ZW$). Other sources, including SM backgrounds producing two or more prompt photons (e.g. $\gamma\gamma+$jets, $W\gamma\gamma$), have small cross sections after applying our selection cuts and are therefore negligible.
 For the $W+$jets, $t\bar{t}+$jets, and $Z+$jets backgrounds, matrix-element samples were generated including up to two additional partons. For processes with an explicit photon, namely $W\gamma$ and $t\bar{t}\gamma$, as well as for single-top production, samples were generated including up to one additional jet. All other backgrounds were generated without extra matrix-element partons. Additional photons may arise from QED radiation in Pythia, while Delphes can misidentify other objects as photons according to its detector model. Both tools were used with their default configurations.

\begin{table}[ht]
\centering
\begin{tabular}{c|c}
\textbf{Requirement} & \textbf{Description} \\
\hline
Leptons & At least one light charged lepton ($e$ or $\mu$) \\
Photons & At least one photon \\
Jets & At least one jet with leading jet satisfying $p_T > 100$~GeV \\
MET & $E_T^\text{miss} > 100$~GeV \\
\end{tabular}
\caption{Event selection criteria employed in the analysis.}
\label{table:selection}
\end{table}
Object identification criteria are outlined in Table~\ref{table:objectid}. These selections follow standard ATLAS Run 2 optimized cuts, following detector performance guidelines. Event preselection follows the criteria summarized in Table~\ref{table:selection}. These cuts are designed to enhance signal sensitivity while reducing the impact of the SM background. We adopt a relatively loose threshold for the missing energy, which can be beneficial in the high-luminosity LHC (HL-LHC) environment by increasing the event yield and potentially reducing statistical uncertainties. As shown in~\cite{Arganda:2024tqo}, even higher thresholds---such as $E_T^\text{miss} > 200$~GeV, compatible with current trigger menus---have only a minimal impact on the overall signal efficiency for this type of final state.

\begin{table}[ht]
\centering
\begin{tabular}{c|c}
\textbf{Low-level features} & \textbf{High-level features} \\
\hline
Leading jet: $p_T^{j_1},\; \eta^{j_1}$ & Hadronic activity: $H_T^{\text{jets}} = \sum p_T^{\text{jets}}$ \\
Leading lepton: $p_T^{\ell_1},\; \eta^{\ell_1}$ & Total transverse energy:  $H_T = \sum (p_T^{\text{jets}} +  p_T^{\tau} + p_T^{e} +  p_T^{\mu} +  p_T^{\gamma})$ \\
Leading photon: $p_T^{\gamma_1},\; \eta^{\gamma_1}$ & Transverse masses: $m_T^{j_1},\; m_T^{\ell_1},\; m_T^{\gamma_1}$ \\
MET: $E_T^\text{miss}$ & $s_T^{1} = p_T^{\ell_1} + p_T^{j_1} + p_T^{\gamma_1}$ \\
Object multiplicities $n_\gamma$, $n_\ell$, $n_j$ & MET significance: $E_T^\text{miss}/\sqrt{H_T}$ \\
\end{tabular}
\caption{List of kinematic variables used as input features for our multivariate analysis.}
\label{table:features}
\end{table}

After applying the baseline event selection, we examined a comprehensive set of variables to describe the final-state kinematics, including basic detector-level variables and more intricate observables enhancing signal-to-background separation. The full list of variables used in the analysis is shown in Table~\ref{table:features}, categorized as low-level and high-level features. These variables constitute the full set of input features for the supervised binary classifier used in our multivariate analysis that we describe next. 

\subsection{Machine Learning Implementation and Statistical Treatment}
\label{sec:ML}
%
To analyze the simulated data, we adopt the strategy outlined in \cite{Arganda:2024tqo}, which is based on the MLL method introduced in \cite{Arganda:2022qzy,Arganda:2022mrd,Arganda:2022zbs}.
This approach leverages \texttt{XGBoost}~\cite{Chen:2016btl} algorithm, a gradient-boosted decision tree method, as its core classifier to distinguish between signal and background events. \texttt{XGBoost} is particularly well-suited for this task due to its ability to handle complex, high-dimensional datasets
and its built-in regularization mechanisms that help prevent overfitting. It has become a standard tool in high-energy physics analyses and has been successfully applied in several ATLAS and CMS studies in several different contexts. 

For the training and validation set, we have generated 200k background events, properly accounting for the weights associated with the different background processes, and 200k signal events, with an equal number of events per BP.
In this way, we ensure that we have a balanced dataset for the training of the machine-learning classifier.
It is also worth noting that combining all BPs into a single dataset may not be optimal when analyzing an individual BP. However, this approach results in a more general model that can be applied across the entire parameter space without the need for retraining.

In addition, we have created a completely independent test set composed of 400k background and signal events, again taking into account the relative weights of the SM processes and an equal number of events per BP.
This set is used to test the final performance of the ML~model~after~training. 

We performed a grid search to find the optimal set of hyperparameters that maximizes the AUC, a standard metric to assess classifier performance, with values ranging between 0.5 and 1. The selected values include a maximum tree depth of 5, a learning rate of 0.01, 2500 estimators, and an L2 regularization term on weights of 0.01. The \texttt{objective} parameter was set to \texttt{binary:logistic} and the evaluation metric to \texttt{logloss}. To avoid overfitting, we used an early stopping of 50 iterations, which ends the training if the performance does not improve when tested over the validation dataset. Importantly, we have found that, for our problem, the \texttt{XGBoost} algorithm is not very sensitive to hyperparameter variations around their optimal values, to random weight initialization, nor to the partition of the train-test-validation datasets. This leads to AUC variations of $\sim0.01$ that do not affect our results significantly and therefore are neglected in the rest of our work.

It is worth noting that, although the input of a classifier can be high-dimensional, its output is one-dimensional and represents the probability of a given observation being a signal event.
Therefore, given a set of events, we can compute both exclusion ($Z = 2$) and discovery ($Z = 5$) sensitivities using the full one-dimensional output of the machine learning model, $o(x)$, in a similar fashion as any other variable.
This is carried out using two different approaches: the Binned Likelihood (BL) framework and an unbinned method called Machine Learned Likelihood (MLL)~\cite{Arganda:2022qzy,Arganda:2022mrd,Arganda:2022zbs}.

In the BL approach, the classifier output $o(x)$ is discretized into a histogram. Then, we determine the expected number of background, $B_{d}$, and signal, $S_{d}$, events in each bin $d$ and compute the likelihood, $\mathcal{L}$, assuming a Poisson distribution for each bin, that is
\begin{equation}
    \mathcal{L}(\mu,s,b) = \prod_{i=1}^{N} \text{Poiss}\big(N_d|\mu S_d + B_d\big) \,,
\end{equation} 
where the signal strength $\mu$ defines the hypothesis we are testing for (discovery corresponds to studying the background-only hypothesis $\mu = 0$).
Then, this likelihood is used to estimate the test statistic, $\tilde{q}_{0}$, as a log-likelihood ratio
\begin{equation}
\tilde{q}_{0} = 
-2 \text{ Ln } \frac{\mathcal{L}(0,s,b)}{\mathcal{L}(\hat{\mu},s,b)} \,,
\end{equation}
with $\hat{\mu}$ is the parameter that maximizes the likelihood. Finally, using the Asimov datasets, described in \cite{Cowan:2010js}, the discovery significance can be obtained through
\begin{equation}
    Z_{\text{BL}}=\text{med }[Z_{0}|1] = \sqrt{\text{med }[\tilde{q}_{0}|1]}=\left[2\sum_{d=1}^{D}\left((S_d+B_d)\text{ Ln}\left(1+\frac{S_d}{B_d}\right)-S_d\right)\right]^{1/2}\,.
    \label{binned-Z}
\end{equation}
Although widely used, this method introduces an inherent information loss due to the binning process, which can smear out subtle features in the output distribution.

In contrast, the MLL method avoids this issue by performing an unbinned fit. In this case, the likelihood function of $N$ independent measurements, each one consisting of a high-dimensional set of observations, $x$, is modeled as
\begin{equation}
    \mathcal{L}(\mu,s,b) = \mathrm{Poiss}(N|\mu S + B) \prod_{i=1}^{N}\left[ \frac{B}{\mu S + B}p_{b}(x_{i}) + \frac{\mu S}{\mu S + B}p_{s}(x_{i})\right] \,,
    \label{eq:mll}
\end{equation}
where the Poisson term describes global information with $S$ and $B$ the total number of signal and background events satisfying $N = S+B$, while $p_{s}(x)=p(x|s)$ and $p_{b}(x)=p(x|b)$ are the signal and background probability density functions (PDFs) for $x$, encoding event-by-event information. Taking advantage of Kernel Density Estimators (KDEs), a non-parametric technique, one can approximate the high-dimensional PDFs with the classifier output PDFs for pure signal and background samples. The resulting KDE-based distributions, $\tilde{p}_s(o(x))$ for signal and $\tilde{p}_b(o(x))$ for background, are then directly inserted into Eq.~\eqref{eq:mll}.

As before, the discovery reach corresponds to the background-only hypothesis ($\mu = 0$), then the log-likelihood ratio test statistic is
\begin{equation}
\tilde{q}_{0} =   \begin{cases}
0 & \rm{if} \, \, \hat{\mu} < 0 \\
-2 \text{ Ln } \frac{\mathcal{L}(0,s,b)}{\mathcal{L}(\hat{\mu},s,b)} =  -2\hat{\mu} S + 2 \sum_{i=1}^{N} \text{ Ln } \left( 1 + \frac{\hat{\mu} S \tilde{p}_s(o(x_i))}{B \tilde{p}_b(o(x_i))}\right)  &  \rm{if} \, \,  \hat{\mu} \geq 0\,.
\end{cases}
  \label{eq:testdiscovery}
\end{equation}
Here, $\hat{\mu}$ has to be estimated numerically by calculating the zero of the partial derivative of Eq.~\eqref{eq:mll} with respect to $\mu$. Finally, the expected discovery significance is defined as
\begin{equation}
    Z_{\text{MLL}}=\text{med }[Z_{0}|1] = \sqrt{\text{med }[\tilde{q}_{0}|1]}\,,
\end{equation}
where we estimate the $\tilde{q}_{0}$ distribution numerically by generating a large set of pseudo-experiments with signal-plus-background events, that is, samples with $\mu'=1$~\footnote{For a full description of the MLL method please refer to \cite{Arganda:2022qzy,Arganda:2022zbs}, while in Refs.~\cite{Arganda:2023qni,Lopez-Fogliani:2024gzj,Arganda:2024tqo} collider and DM direct detection experiments studies applying this approach can be found.}.

\subsection{Results}
\label{sec:results}

\begin{figure}
    \centering
    \includegraphics[width=0.4\linewidth]{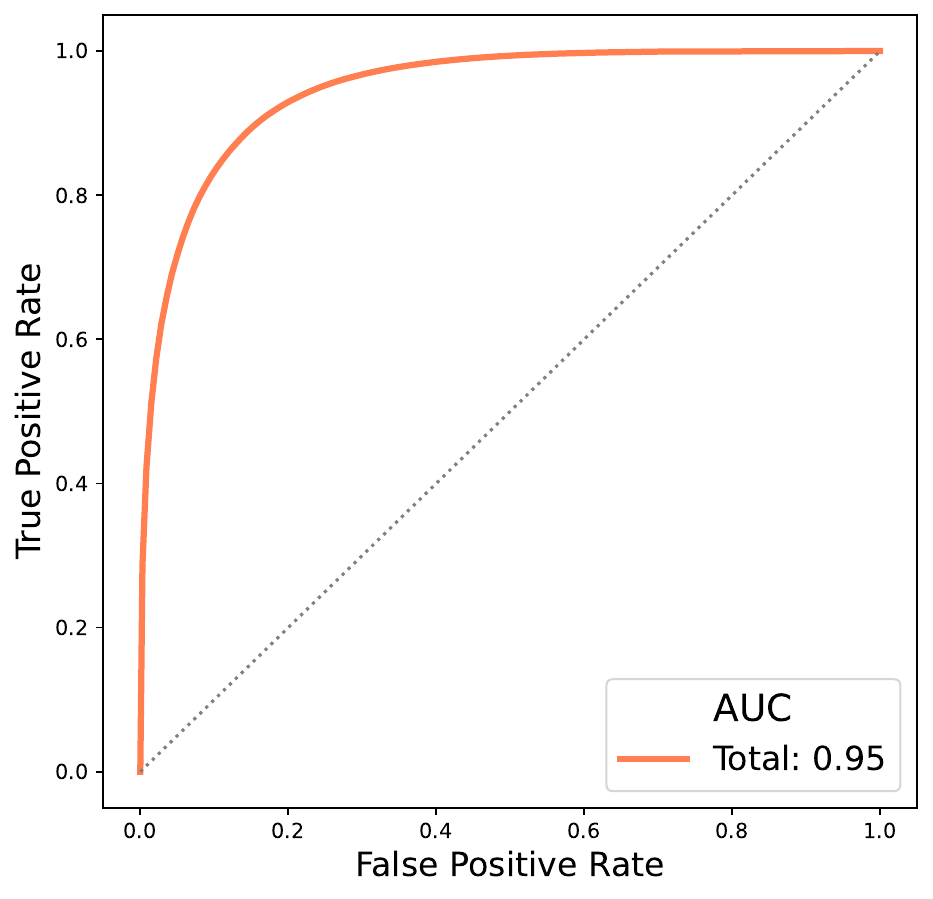}
    ~~~~~~
    \includegraphics[width=0.5\linewidth]{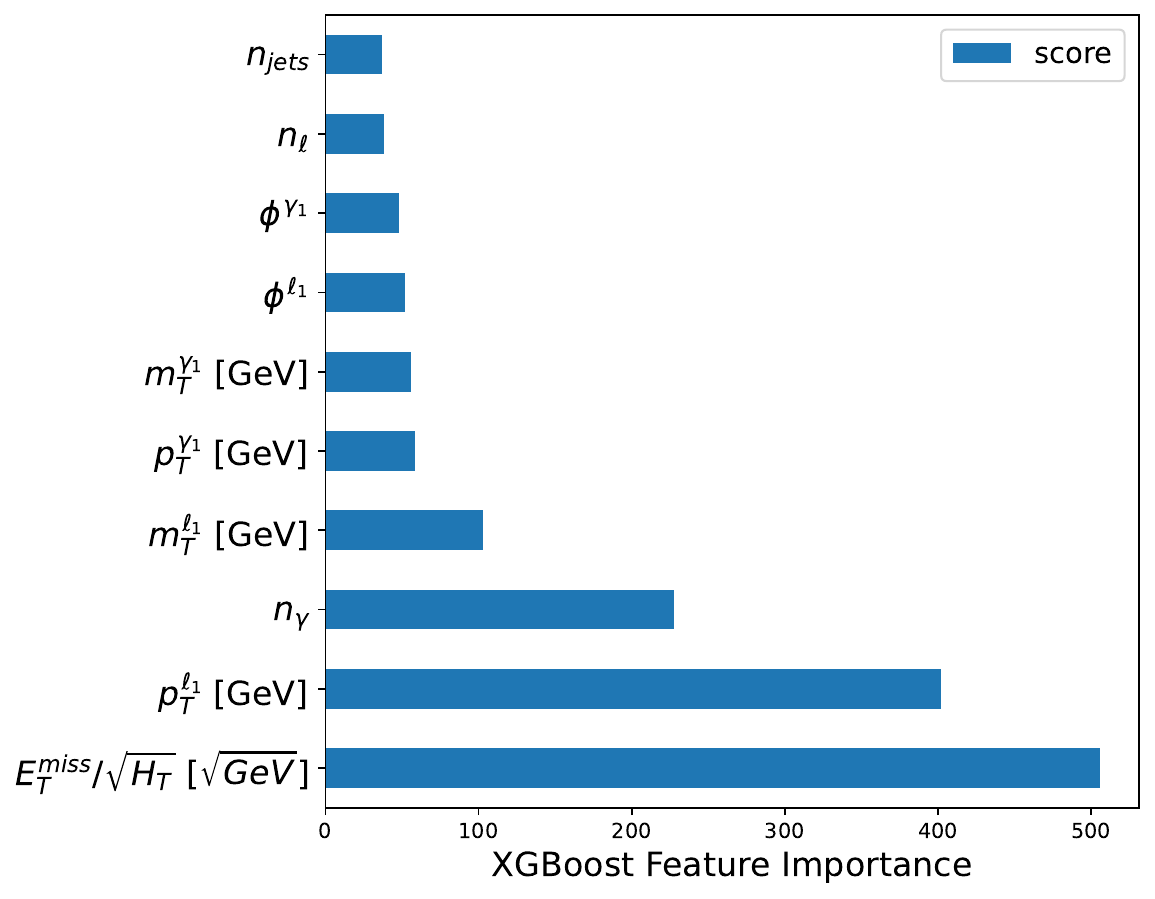}
    \caption{Left panel: ROC curve and AUC performance metric of the XGBoost classifier trained using all the BPs as signal. Right panel: feature importance score (gain metric) for the same XGBoost classifier.}
    \label{fig:roc}
\end{figure}

In this section, we present the main results of the analysis. We begin by showing the performance of the binary classifier in Figure~\ref{fig:roc}. The left panel displays the Receiver Operating Characteristic (ROC) curve, which provides a visual representation of signal efficiency and background rejection across different decision thresholds. We can see that our \texttt{XGBoost} algorithm performs very well in distinguishing between background and signal events, with an AUC (area under the ROC curve) of 0.95.

In the right panel of Figure~\ref{fig:roc}, we show the 10 most relevant features for classifying signal and background events. This is determined by measuring the contribution of a feature on the discrimination task (gain metric), with a higher value indicating a higher impact. The most relevant features are the missing transverse energy significance (higher values indicate the presence of invisible particles escaping the detector), leading lepton characteristics (momentum and transverse mass), and photon features (object multiplicity, and momentum and transverse mass of the leading one).

\begin{figure}
    \centering
    \includegraphics[width=1\textwidth]{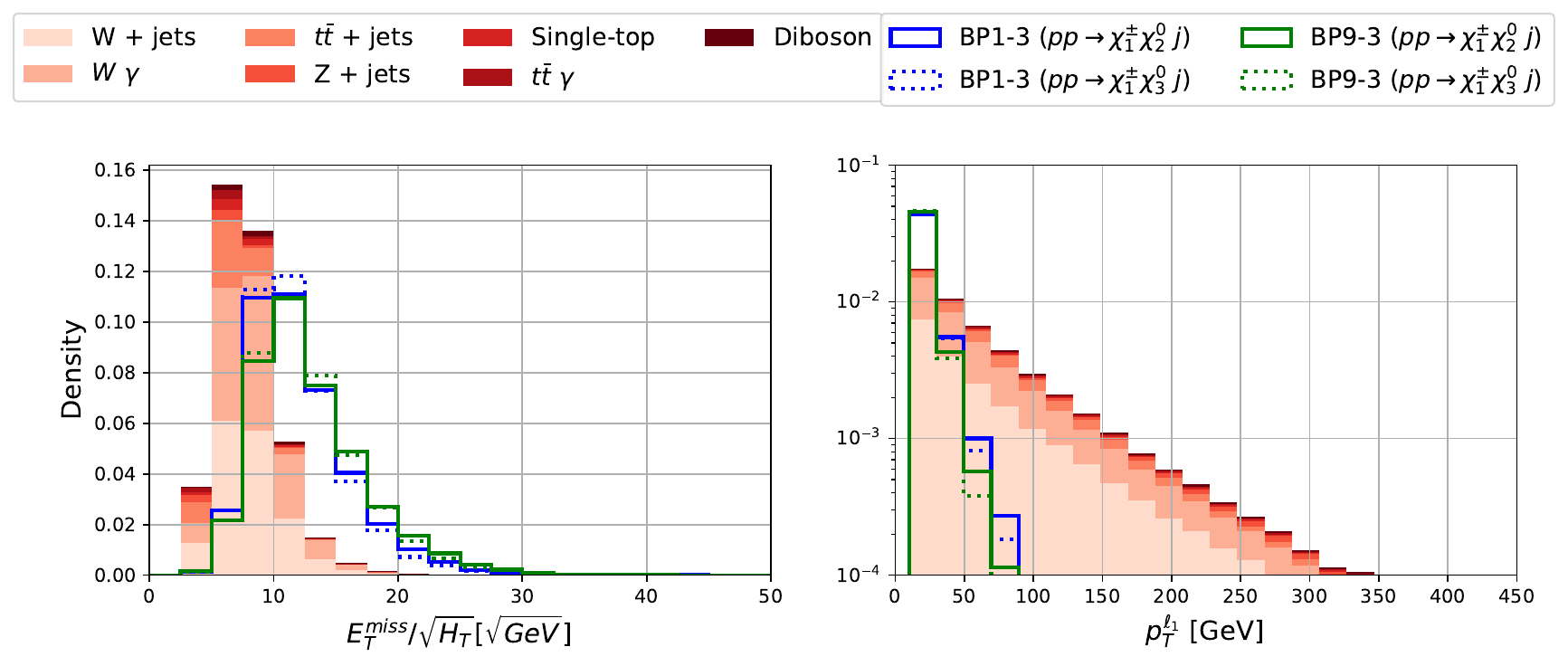}
    \includegraphics[width=0.96\textwidth]{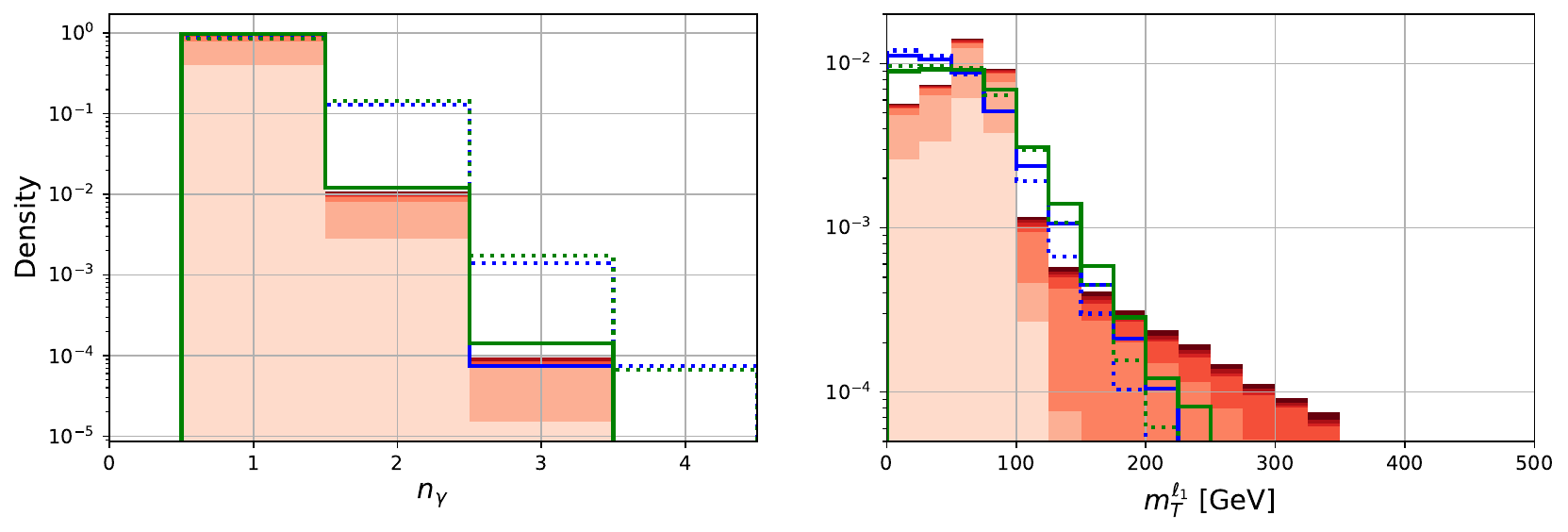}
    \includegraphics[width=0.96\textwidth]{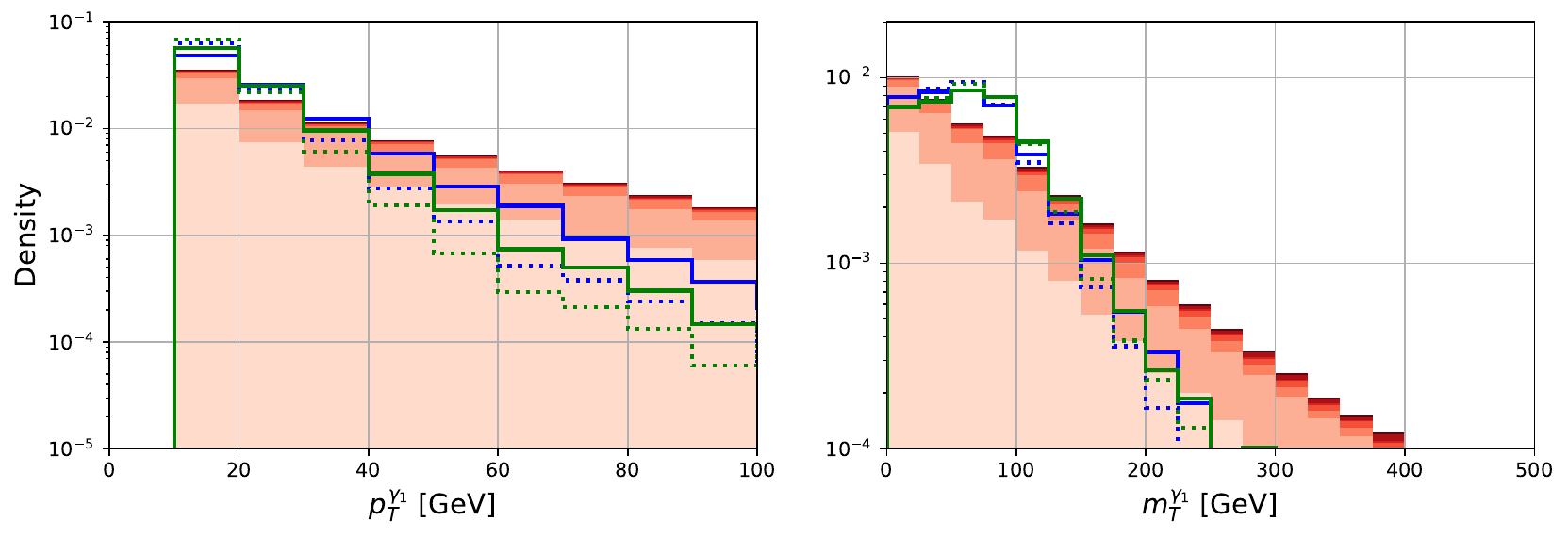}
    \caption{Distributions of the six final state kinematic variables showing the most discrimination between signal and background: missing transverse energy significance, $E_T^\text{miss}/\sqrt{H_T}$ (top-left panel); total transverse momentum of the leading lepton, $p_T^{\ell_1}$ (top-right panel); number of photons, $n_{\gamma}$ (center-left panel); transverse mass of the leading lepton, $m_T^{\ell_1}$ (center-right panel); transverse momentum of the leading photon, $p_T^{\gamma_1}$ (bottom-left panel); and transverse mass of the leading photon, $m_T^{\gamma_1}$ (bottom-right panel). We display two representative benchmark points: BP1-3 with low masses ($\tilde\chi_{1,2,3}^0 \sim 145-165$ GeV), and BP9-3 with higher masses ($\tilde\chi_{1,2,3}^0 \sim 235-250$ GeV). Each production channel is depicted separately: $pp \rightarrow \tilde\chi^{\pm}_1 \tilde\chi^0_2 j$ (solid curves) and $pp \rightarrow \tilde\chi^{\pm}_1 \tilde\chi^0_3 j$ (dotted curves) }
    \label{fig:relevant-variables}
\end{figure}

To improve interpretability, in Figure~\ref{fig:relevant-variables}, we present the one-dimensional distributions of the six most important features. We show the SM contribution per channel with different tones of red as stack histograms. To illustrate the signal behavior, we selected two representative benchmark points (see Table~\ref{table:bpstext}): BP1-3 in blue with low masses ($\tilde\chi_{1,2,3}^0 \sim 145-165$ GeV), and BP9-3 in green with higher masses ($\tilde\chi_{1,2,3}^0 \sim 235-250$ GeV). For these BPs the neutralino DM can reproduce the majority of the measured abundance with $\Omega_{\tilde\chi^0_1}h^2=0.1$ and $0.07$, and as we will see, show a promising detection prospect in our collider analysis with significances of $Z>5$ and $Z>2$, respectively. Finally, we also depict the contribution of each production channel independently: $pp \rightarrow \tilde\chi^{\pm}_1 \tilde\chi^0_2 j$ with solid curves and $pp \rightarrow \tilde\chi^{\pm}_1 \tilde\chi^0_3 j$ with dotted curves.

\begin{table}[t]
\centering
\begin{threeparttable}
\small
\setlength{\tabcolsep}{6pt}
\renewcommand{\arraystretch}{1.1}
\begin{tabularx}{0.55\textwidth}{@{}l *{2}{>{\centering\arraybackslash}X}@{}}
\toprule
\textbf{BP} & \textbf{BP1–3} & \textbf{BP9–3} \\
\midrule
$\mntrlone$ [GeV]                       & 147.5 & 235.1 \\
$\mntrltwo$ [GeV]                       & 158.5 & 245.0 \\
$\mntrlthree$ [GeV]                       & 164.8 & 251.7 \\
$\mcharone$ [GeV]                           & 161.8             & 248.9             \\
BR$(\tilde\chi^0_2 \!\to\! \tilde\chi^0_1\gamma)$   & 0.73              & 0.82              \\
BR$(\tilde\chi^0_3 \!\to\! \tilde\chi^0_2\gamma)$   & 0.89              & 0.91              \\
BR$(\tilde\chi^\pm_1 \!\to\! \tilde\chi^0_1 W^\pm)$ & 0.96              & 0.92              \\
$\sigma(pp\!\to\!\tilde\chi^\pm_1\tilde\chi^0_2 j)$ [fb]     & 105.1             & 28.4              \\
$\sigma(pp\!\to\!\tilde\chi^\pm_1\tilde\chi^0_3 j)$ [fb]     & 99.1              & 27.1              \\
$\sigma_\text{DD}^{\mathrm{SI}}$ [cm$^{2}$]                 & $1.3\times 10^{-48}$ & $4.2\times 10^{-48}$ \\
$\Omega_{\tilde\chi^0_1} h^2$                          & 0.10              & 0.07              \\
$Z_\text{BL}$                                               & $6.29$                & $2.84$                \\
$Z_\text{MLL}$                                              & $6.67$               & $3.80$                \\
\bottomrule
\end{tabularx}
\begin{tablenotes}[flushleft]
\footnotesize
\item Cross sections include one jet with $p_T>100~\text{GeV}$ and $|\eta_j|<2.5$.
\end{tablenotes}
\caption{Benchmark points used in Figure~\ref{fig:relevant-variables}.}
\label{table:bpstext}
\end{threeparttable}
\end{table}

In the top-left panel of Figure~\ref{fig:relevant-variables}, we present the most discriminant feature, the missing transverse energy significance, $E_T^\text{miss}/\sqrt{H_T}$. Large values of this variable indicate that the observed $E_T^\text{miss}$ is unlikely to be explained by momentum resolution effects, implying the presence of undetected objects. For background events, the contribution comes from neutrinos produced in $W$ decays. For signal events, $E_T^\text{miss}$ is dominated by the contribution of the two $O(100)$ GeV lightest neutralinos in the final state that escape the experiment undetected, resulting in a harder MET significance distribution than in the SM background case. As expected, the BP9-3 distribution is slightly skewed towards larger values compared to BP1-3, due to heavier neutralinos.


The transverse momentum of the leading lepton, $p_T^{\ell_1}$, is the second most important feature (shown in the top-right panel of Figure 3). For the SM backgrounds involving a $W$ boson, especially the dominant ones $W+$jets and $W\gamma$, leptons are mostly produced by the decay of on-shell $W$ bosons. \footnote{We have verified that, after the selection cuts, the contribution of diagrams involving off-shell $W$ is small and does not significantly alter the event yield or the distributions of the most relevant features.} In contrast, for signal events, leptons arise from off-shell $W$-boson decays produced in the decay of the lightest chargino to the lightest neutralino, $\tilde\chi_1^\pm \rightarrow \tilde\chi_1^0 \ell \nu_\ell$, which are very close in mass due to the compressed spectra considered in this work. Therefore, the available phase space is smaller, resulting in a softer lepton $p_T{\ell_1}$ distribution.

The third most important feature corresponds to the number of photons in the final state, $n_{\gamma}$, and is shown in the center-left panel of Figure~\ref{fig:relevant-variables}. Interestingly, there is not a significant difference between the signal production channel $pp \rightarrow \tilde\chi^{\pm}_1 \tilde\chi^0_2 j$ (solid curves) and the SM background expected number of photons. However, we can clearly see the importance of the production channel $pp \rightarrow \tilde\chi^{\pm}_1 \tilde\chi^0_3 j$ (dotted curves) and the key role that it plays in the discrimination procedure, due to a higher production of photons thanks to a longer decay chain.

The transverse mass of the leading lepton, $m_T^{\ell_1}$, the fourth most-important feature, is shown in the center-right panel of Figure~\ref{fig:relevant-variables}. Given that the SM contribution is dominated by $W+$jets and $W\gamma$, the background distribution peaks around the $W$-boson mass. This is expected, since the lepton and the neutrino, the only source of $E_T^{\rm miss}$, come from the on-shell $W$ decays. On the other hand, the signal presents a broader distribution since the $E_T^\text{miss}$ accounts not only for neutrinos but for two neutralinos LSP.

The fifth most important feature, the transverse momentum of the leading photon, $p_T^{\gamma_1}$, is shown in the bottom-left panel of Figure~\ref{fig:relevant-variables}. As expected, the signal distribution peaks at lower values than the background one. As in the lepton case, this is due to the compressed mass spectra of the light electroweakinos, whose decays ($\tilde\chi_{2(3)}^0 \rightarrow \tilde\chi_{1(2)}^0 \gamma$) produce not too energetic photons.

To conclude the discussion of the most important features for classification, the transverse mass of the leading photon, $m_T^{\gamma_1}$, is shown in the bottom-right panel of Figure~\ref{fig:relevant-variables}. We can see that the signal distribution peaks at $\sim 100$ GeV, similar value to the event selection criterion for the leading jet ($p_T > 100$ GeV). This indicates that this feature reconstructs the energy of the neutralino-chargino system that recoils against the ISR jet, since the lepton produced in the chargino decay chain is soft (see top-right panel of Figure~\ref{fig:relevant-variables}). 

\begin{figure}[t!]
    \centering
    \includegraphics[width=0.85\textwidth]{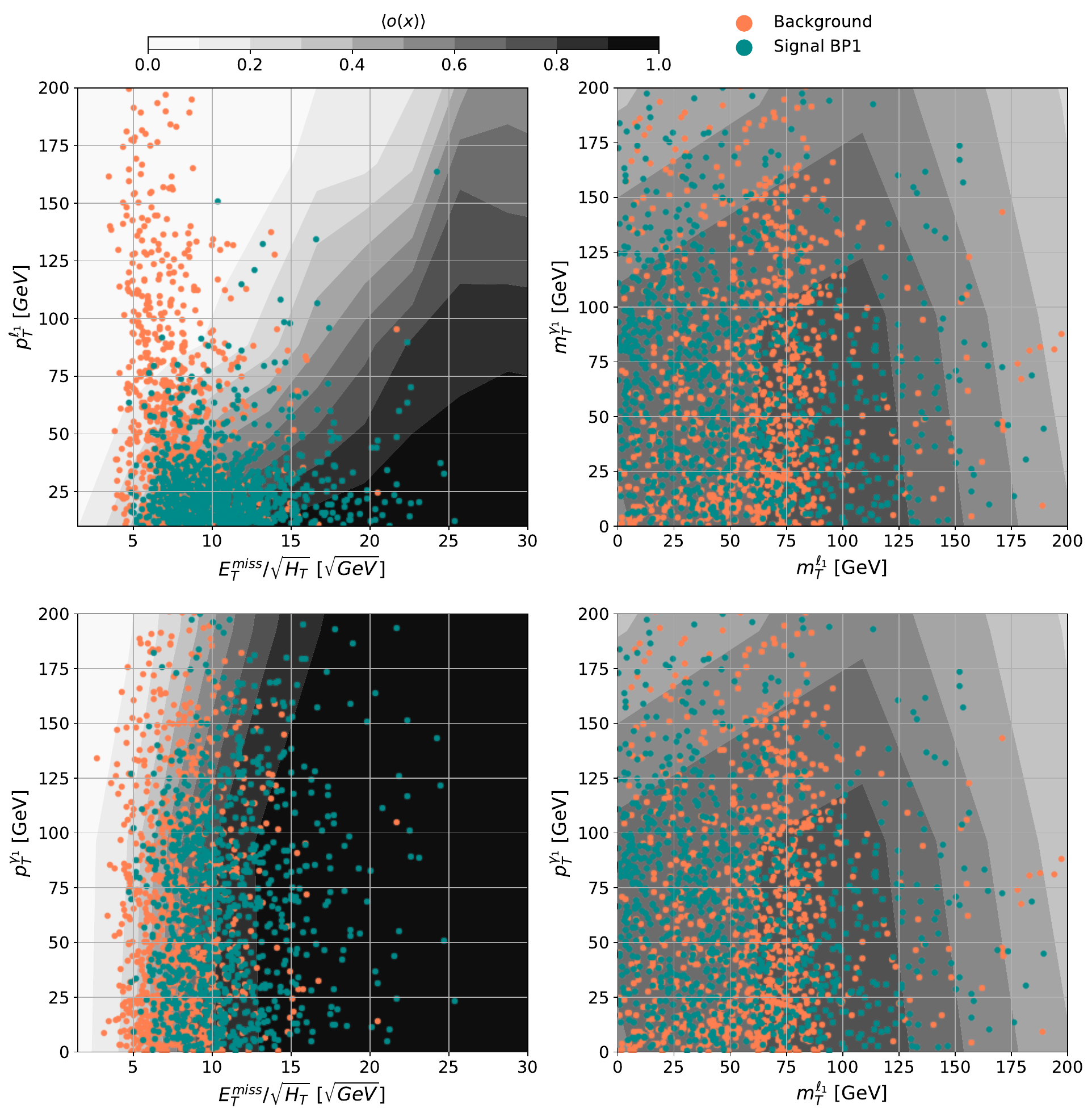}
    \caption{2D distributions between some of the most important features in the classification task. Green (orange) points correspond to signal (background) events. The gray scale shows the average ML output, $\left< o(x) \right>$ considering the points inside each grid-like 2D bin.}
    \label{fig:correlations}
\end{figure}

One way to visualize the classification made by the ML algorithm is to analyze its result in 2D projections. In Figure~\ref{fig:correlations} we show the 2D distributions between some of the most important features, with green (orange) points corresponding to signal (background) events. In gray scale, we show the behavior of the ML output, $o(x)$, which is computed by dividing the 2D space into a grid and averaging the ML output, $\left< o(x) \right>$, in each bin. A lighter tone indicates that $\left< o(x) \right> \rightarrow 0$, or in other words, an event in that region has a higher probability of being classified as background-like, while darker tones indicate $\left< o(x) \right> \rightarrow 1$, a higher probability of determining that the event is signal-like. In Figure~\ref{fig:correlations}, we can see that the ML contours describe the non-trivial correlation better than one could achieve with rectangular cuts in each variable. For example, from the 1D distributions (first row of Figure~\ref{fig:relevant-variables}) one could propose a signal enriched region with $E_T^\text{miss} / \sqrt{H_T} > 10 \, \text{GeV}^{1/2}$ and $p_T^{\ell_1} < 25$ GeV, which coincides with the darker ML contours in the top-left panel of Figure~\ref{fig:correlations}, indicating signal-like events. However, the same dark tone areas extend beyond the very simple rectangular signal-enriched region and, in principle, could provide the same level of information for more events. Moreover, the employed ML-based methods do not use a signal-enriched region at all but consider the entire parameter space in their statistical treatments, taking advantage of easy-to-classify events as well as the tails of the distributions. For other 2D projections, like the one shown in the bottom-left panel of Figure~\ref{fig:correlations} the ML predictions describe contours similar to those that one would expect following rectangular cuts in $E_T^\text{miss} / \sqrt{H_T}$, while presenting only a mild dependence on $p_T^{\gamma_1}$. 
Regarding the photon features, $m_T^{\gamma_1}$ and $p_T^{\gamma_1}$, we have found that ML contours describe non-trivial correlations with lepton features like $m_T^{\ell_1}$ (right column of Figure~\ref{fig:correlations}), which are very hard to determine by eye from the 1D or the 2D distributions.

\begin{figure}[t]
    \centering
    \includegraphics[width=0.85\textwidth]{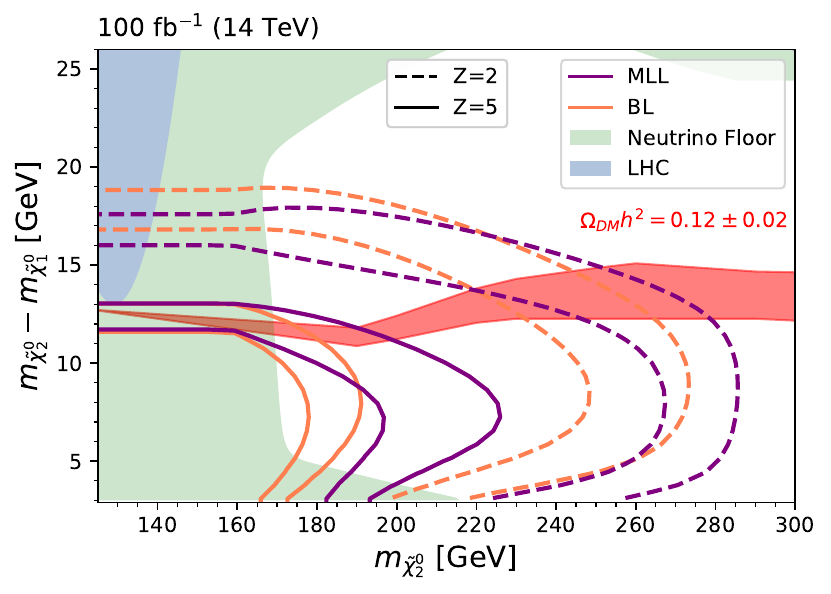}
    \caption{Projected discovery significance $Z$ = 2 (dashed) and $Z$ = 5 (solid) contour lines in the [$m_{\tilde\chi_2^0}$, $m_{\tilde\chi_2^0}-m_{\tilde\chi_1^0}$] plane applying the Binned-Likelihood approach (BL, orange), and the Machine-Learned Likelihood method (MLL, violet). }
    \label{fig:contour_plots}
\end{figure}

The projected LHC sensitivity is presented in Figure~\ref{fig:contour_plots}. For each BP, we generated 2k pseudo-experiments, evaluated the test statistic, and computed its significance including its statistical uncertainty from the variance of the resulting test statistic distribution. In dashed curves, we show the exclusion reach ($Z=2$), while in solid curves we indicate the discovery limit ($Z=5$). These curves correspond to the contour levels computed after interpolating the significance values obtained for each BP. The region between the same-style curves represents the statistical uncertainty, $1\sigma_\text{stat}$, considering a conservative approach where we only allow a weaker sensitivity reach. Both the binned (BL) and unbinned (MLL) methods yield similar results, although the MLL strategy extends the electroweakino masses that can be explored by $\sim10\%$. In summary, using $100~\mathrm{fb}^{-1}$ integrated luminosity data at $14~\mathrm{TeV}$, we achieve a $5\sigma$ discovery reach for higgsino masses  $m_{\tilde\chi_2^0}\!\lesssim225~\mathrm{GeV}$ with $m_{\tilde\chi_2^0}-m_{\tilde\chi_1^0}\!\lesssim\!12~\mathrm{GeV}$, and a $2\sigma$ exclusion up to $285~\mathrm{GeV}$ with $m_{\tilde\chi_2^0}-m_{\tilde\chi_1^0}\!\lesssim\!20~\mathrm{GeV}$.

In order to quantify the improvement obtained by the MLL strategy, we performed a simple cut-based analysis taking into account the two most discriminant variables, namely the MET significance and $p_{t}^{\ell_{1}}$. We found that the MLL strategy yields significances approximately 4.5 times larger than the simple cut-based analysis. This is in agreement with the results found in Ref.~\cite{Arganda:2024tqo}, where the advantages of the MLL method over other methods were shown in a similar setup.

It is interesting to note that for a fixed value of $m_{\tilde\chi_2^0}$, the sensitivity increases for smaller mass splittings, up to $m_{\tilde\chi_2^0}-m_{\tilde\chi_1^0} \sim 7$ GeV. This is because the branching ratio to photons increases for lower mass splittings, while the production cross sections are independent (see Tables~\ref{table:bps2} and \ref{table:bps3}); therefore, we expect more signal events. However, for $m_{\tilde\chi_2^0}-m_{\tilde\chi_1^0}\lesssim7$ GeV, the significance starts to decrease again. In this region, the photon produced in the channel $pp \rightarrow \tilde\chi^{\pm}_1 \tilde\chi^0_2 j$ can be very soft and may not satisfy our detection criteria. This fact highlights the importance of the production channel $pp \rightarrow \tilde\chi^{\pm}_1 \tilde\chi^0_3 j$: it generates extra photons (the third most-important feature for classification as can be seen in the center-left panel of Figure~\ref{fig:relevant-variables}), if $m_{\tilde\chi_2^0}-m_{\tilde\chi_1^0}\lesssim7$ GeV the most energetic photon is produced by the decay $\tilde\chi_3^0 \rightarrow \tilde\chi_2^0 \gamma$ since $m_{\tilde\chi_3^0} - m_{\tilde\chi_2^0}\sim 7$ GeV, and finally, this channel dominates the signal yield (its relative weight after the event selection criteria cuts increases for lower mass splittings $m_{\tilde\chi_2^0}-m_{\tilde\chi_1^0}$). Note that even for electroweakinos with mass splittings below 10 GeV (the lower limit of the photon identification criteria) we have sensitivity due to the fact that the neutralino-chargino system is boosted against a high-$p_T$ ISR jet. A looser lower limit on the photon $p_T$ could be useful in extending the parameter space that could be explored to even lower mass splittings. However, the region close to the ultra-compressed mass scenario would require a dedicated analysis outside the scope of this work.

Following the procedure detailed in Ref.~\cite{Arganda:2023qni}, we can evaluate the impact of the systematic uncertainties in the ML analysis. The estimation considers variations in the most important features for the ML discrimination, and creates a new data set by shifting the values of the mentioned features in all events. Then, the computation of the significance is repeated using the modified data set. Finally, the uncertainty is identified as the variation in significance, discarding any artificial increase of performance. In our case, the most important feature for ML discrimination is $E_T^{miss}/\sqrt{H_T}$ (see the right panel of Figure~\ref{fig:roc}) with a $5\%$ uncertainty according to Ref.~\cite{CMS:2019ctu}, a CMS performance analysis with final states and selection cuts similar to the ones considered in this work. The impact of the systematic uncertainty can be estimated to be smaller than the statistical errors by a factor of $\sim2$ for the unbinned MLL method, and slightly smaller for the binned one. Based on this analysis, we expect that the final error to be dominated by statistical uncertainties. Therefore, we do not include the systematic estimation in our results, and leave the precise assessment of all errors and their combination to the experimental collaborations.

In Figure~\ref{fig:contour_plots} we also show in gray the parameter space that is excluded by recasting current LHC constraints with \texttt{CheckMATE}. As mentioned previously, the resulting bounds of analyses that target electroweakino production with final states containing multiple leptons plus $E_T^\text{miss}$~\cite{ATLAS:2019lng,CMS:2021cox} are significantly weaker in our model due to the suppression of effective branching ratios into leptonic final states. For intermediate mass splittings, the excluded region extends up to $m_{\tilde\chi_2^0}\sim 130-140$ GeV, while with our proposed signal we could explore up to $m_{\tilde\chi_2^0}\sim 280$ GeV.

With a red band we demarcate the parameter space that saturates the observed DM abundance. Points below this band (with a lower degree of compression, $\epsilon$) produce DM underabundance, whereas points above (with higher $\epsilon$) would result in DM overabundance. Since the inclusion of additional SM extensions, such as late-time injection of entropy during the Big Bang, could make these points viable without affecting collider signatures, we do not exclude them to be as independent as possible from early universe processes. Then, to compute direct detection constraints, we rescale the value of $\sigma^\text{SI}_\text{DD}$ shown in Tables~\ref{table:bps2} and \ref{table:bps3} and define an effective SI cross-section as:
\begin{equation}
\sigma^\text{SI-eff}_\text{DD} = \sigma^\text{SI}_\text{DD} \, \text{min}\left[ 1 \,, \, \frac{\Omega_{\tilde\chi_1^0} h^2}{0.12} \right].
\end{equation}
In this way, we assume that the neutralino LSP of our model represents the highest percentage of the current DM budget possible. Points that result in LSP overabundance are rescaled to saturate the measured DM relic density, i.e. $\sigma^\text{SI-eff}_\text{DD} = \sigma^\text{SI}_\text{DD} \,$. On the other hand, points with LSP under abundance, which implies a lower local LSP relic density and therefore a lower direct detection signal generated by the LSP, have their SI cross section reweighted by the factor $\Omega_{\tilde\chi_1^0} h^2 / 0.12 < 1$.
Due to our choice of parameters near the blind-spot condition, especially $M_1 \simeq 500$ GeV (see Section~\ref{constraints}), all BPs satisfy the latest DM direct detection constraints. 

The green region in Figure~\ref{fig:contour_plots} denotes the so-called neutrino floor~\cite{OHare:2021utq}, that is, the region where the sensitivity of a direct DM experiment is limited by the irreducible background from coherent elastic neutrino-nucleus scattering. Its unusual shape can be explained by looking at the behavior of $\sigma^\text{SI-eff}_\text{DD}$. Although for a fixed $m_{\tilde\chi_2^0}$ the value of $\sigma^\text{SI}_\text{DD}$ increases for smaller mass splittings, the relic density decreases (see Tables~\ref{table:bps2} and \ref{table:bps3}), resulting in $\sigma^\text{SI-eff}_\text{DD}$ with a turning point when $\Omega_{\tilde\chi_1^0}h^2 < 0.12$. Next generation facilities like DARWIN/XLZD~\cite{DARWIN:2016hyl,Baudis:2024jnk} and PandaX-xT~\cite{PANDA-X:2024dlo} are planning to explore $O(100)$ GeV WIMPs with sensitivities to cross sections that would reach the neutrino floor considering their full exposure, i.e. the white region in Figure~\ref{fig:contour_plots}. In contrast, we can see that the LHC has a discovery potential ($Z>5$) with a near future luminosity of 100 fb$^{-1}$ for the very interesting scenario of compressed electroweakinos with cross sections even below the neutrino floor (mass splittings up to 20 GeV and $m_{\tilde\chi_2^0}\sim170$ GeV).

\begin{figure}
    \centering
    \includegraphics[width=0.85\linewidth]{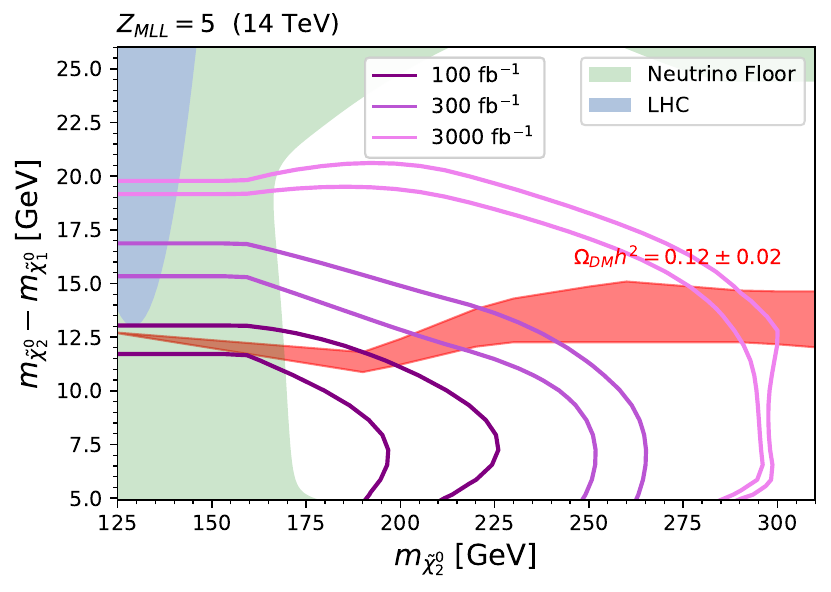}
    \caption{Expected discovery regions ($Z_\text{MLL} > 5$) for future higher luminosities. In purple, violet and magenta we show the results corresponding to $100$ fb$^{-1}$, $300$ fb$^{-1}$ and $3000$ fb$^{-1}$ respectively.}
    \label{fig:Zhighlumi}
\end{figure}

Taking into account that the HL-LHC is expected to reach a luminosity of $3000$ fb$^{-1}$ by the end of its operation, we perform a forecast of the discovery regions ($Z_\text{MLL} > 5$) by re-scaling our significances to higher luminosities. We present the results in Figure~\ref{fig:Zhighlumi} where we show the corresponding regions in purple ($100$ fb$^{-1}$), violet ($300$ fb$^{-1}$) and magenta ($3000$ fb$^{-1}$). The rest of the curves shown in Figure~\ref{fig:contour_plots} follow a similar trend. For the expected full luminosity, electroweakinos with mass splittings up to 20 GeV and $m_{\tilde\chi_2^0}\sim300$ GeV could be discovered, including significant regions of the parameter space that reproduce the observed DM abundance.

Finally, we would like to mention that the direct detection signal, and therefore its allowed parameter space and the location of the neutrino floor, depends on the chosen parameter values in our analysis, especially on $M_1$ as pointed out before. Remarkably, the collider reach is not significantly altered, and in this study we obtain an impressive potential coverage of the parameter space. This not only highlights the power of collider searches, and in particular of the proposed novel signal channel, to probe DM candidates that remain hidden from current and future direct-detection experiments, but also stresses the importance of their complementarity to shed light on the parameter values and to disentangle between models. Therefore, we believe that a more in-depth and dedicated analysis of uncertainties and background modeling that affect this exciting channel is worth pursuing by the LHC experimental collaborations.
%

\section{Conclusions}
\label{sec:conclusions}

As the LHC increases its luminosity, the exploration of signatures with previously prohibited small cross sections becomes an exciting possibility. In that context, ML tools are vital in the search for new physics to overcome the increasingly overwhelming SM background. In this work, we applied ML techniques on the search for DM in a compressed mass scenario with a novel signature that presents one or two soft photons, a soft lepton, and missing energy in the final state.

Specifically, we studied a singlino-dominated LSP as a WIMP DM candidate within the $Z_3$ symmetric NMSSM. Due to its low annihilation rate, a pure singlino leads to an overabundance of relic DM. This issue can be alleviated through co-annihilation with nearby electroweakinos, which in our case are higgsino-like neutralinos close in mass with the LSP, i.e. a compressed electroweakino spectrum. This theoretical framework is especially compelling as it naturally gives rise to so-called \textit{blind spots} where the DM-nucleus cross section can be strongly suppressed, satisfying the stringent DM direct detection results. The blind spot condition is governed by the relative sign between $\mu_\text{eff}$ and $M_1$, and is sensitive to the absolute value of $M_1$, which we take as $\sim500$ GeV. Importantly, the collider signature considered in this work is not significantly modified by the particular~tuning~of~$M_1$.

From a collider perspective, this scenario is particularly attractive because of the pair production of a chargino ($\tilde\chi_1^{\pm}$) plus a neutralino ($\tilde\chi_2^0$ or $\tilde\chi_3^0$), both higgsino-like, and enhanced radiative decay modes of higgsinos into the singlino-dominated LSP and a photon. Although they have small production cross sections, the large luminosities expected at the LHC in the coming years make the exploration of this a novel signature a possibility that has not been explored extensively yet. 

The small mass splittings between the LSP and its co-annihilation partners produce an enhancement of the radiative decays to photons with respect to the well-known channels with di-lepton and tri-lepton in the final state. However, it also implies soft final states, including leptons and photons, which poses a significant challenge for traditional search methodologies. To address these challenges, we implemented two ML-based analyses that leverage the full correlation between kinematic variables, a capability often beyond cut-based approaches. Focusing on LHC proton-proton collisions at $\sqrt{s} = 14$ TeV with an integrated luminosity of $100$ fb$^{-1}$, we employed both a Binned-Likelihood strategy and an unbinned method called Machine-Learned Likelihood. Both results are compatible, showing consistency, and demonstrate the potential of ML techniques to explore large regions of the parameter space. Specifically, our analysis projects a $5\sigma$ discovery reach for $m_{\tilde\chi_2^0}$ up to $225$ GeV with mass splittings ($m_{\tilde\chi_2^0} - m_{\tilde\chi_1^0}$) of approximately $12$ GeV, and a $2\sigma$ exclusion limit extending up to $285$ GeV with mass splittings around $20$ GeV. Remarkably, the projected sensitivity regions cover parameter space points that are inaccessible to current or future DM direct detection experiments, even falling below the so-called neutrino floor, illustrating the key complementarity of LHC searches and DM direct detection approaches.

Considering the promising results in this article, we strongly encourage the LHC experimental collaborations to perform a more in-depth and detailed analysis. This novel search channel with photons in the final state has the potential to complement the traditional and ongoing di-lepton and tri-lepton plus missing energy searches to explore compressed mass scenarios. Moreover, separated and dedicated analyses considering only one or only two photons in the final state could be considered to gain insight into particular production channels, $pp \rightarrow \tilde\chi^{\pm}_1 \tilde\chi^0_2 j$ and $pp \rightarrow \tilde\chi^{\pm}_1 \tilde\chi^0_3 j$, respectively.
%

\vspace{-3.0mm}
\paragraph{\small Acknowledgments.}
%
%
{\small
This work is partially supported by the Spanish Research Agency (Agencia Estatal de Investigaci\'on) through the grants IFT Centro de Excelencia Severo Ochoa No CEX2020-001007-S (EA, AP, RMSS), PID2021-124704NB-I00 (EA, RMSS), PID2021-125331NB-I00 (AP), CNS2023-144536 (RMSS), and RYC-2017-22986 (RMSS), funded by MCIN/AEI/10.13039/501100011033, and by the Comunidad Aut\'onoma de Madrid through the grant SI2/PBG/2020-00005 (AP).
MdlR is supported by the Next Generation EU program, in the context of the National Recovery and Resilience Plan, Investment PE1 – Project FAIR ``Future Artificial Intelligence Research''.
SR is supported by the U.S.~Department of Energy under contracts No.\ DEAC02-06CH11357 at the Argonne National Laboratory. SR would like to thank the University of Chicago, Fermilab, the Aspen Center for Physics, and the Perimeter Institute for Theoretical Physics, where a significant part of this work was carried out.
The work of CW\ at the University of Chicago has been supported by the DOE grant DE-SC0013642. CW\ would like to thank the Aspen Center for Physics, which is supported by National Science Foundation grant No.~PHY-1607611, where part of this work~has~been~done.
}

\newpage
\appendix

\section{Benchmark Points}
\label{sec:BPs}
Benchmark points that are used for our ML-based analysis are listed in the following two Tables~\ref{table:bps2} and \ref{table:bps3}.

\begin{table}[htbp]
\centering
\begingroup
\scriptsize
\setlength{\tabcolsep}{5pt}
\renewcommand{\arraystretch}{1.05}
\setlength{\extrarowheight}{-0.7pt}
\setlength{\aboverulesep}{0.25ex}
\setlength{\belowrulesep}{0.25ex}
\setlength{\abovetopsep}{0.3ex}
\setlength{\belowbottomsep}{0.3ex}
\renewcommand\cellgape{\relax}

\begin{adjustbox}{max width=\textwidth}
\begin{tabular}{@{}c|c c c c c c c c@{}} 
  \toprule
  \textbf{BP} &
  \makecell{$m_{\tilde\chi_1^0,\tilde\chi_2^0,\tilde\chi_3^0}$\\[-2pt]{[GeV]}} &
  \makecell{$m_{\tilde\chi_1^{\pm}}$\\[-2pt]{[GeV]}} &
  \makecell{BR$(\tilde\chi_2^0\!\to\!\tilde\chi_1^0\gamma)$,\\[-2pt]
           BR$(\tilde\chi_3^0\!\to\!\tilde\chi_2^0\gamma)$,\\[-2pt]
           BR$(\tilde\chi_1^{\pm}\!\to\!\tilde\chi_1^0 W^{\pm*})$} &
  \makecell{$\sigma(pp\!\to\!\tilde\chi_1^\pm \tilde\chi_2^0 j)$,\\[-2pt]
           $\sigma(pp\!\to\!\tilde\chi_1^\pm \tilde\chi_3^0 j)$\\[-2pt]
           $(p_T(j)\!>\!100\,\mathrm{GeV},~|\eta_j|\!<\!2.5)$\\[-2pt][fb]} &
  $\sigma_\text{DD}^\text{SI}$ [cm$^{2}$] &
  $\Omega_{\tilde\chi_1^0} h^2$ & $Z_\text{BL}$ & $Z_\text{MLL}$ \\
  \midrule
    \bf 1-1 & 133.6, 158.5, 164.8 & 161.8 & 0.28, 0.79, 0.98 & 105.1, 99.1 & $2.5 \times 10^{-49}$ & 7.5 & $0.83$ & $0.23$\\
    \bf 1-2 & 141, 158.5, 164.8 & 161.8 & 0.47, 0.87, 0.98 & 105.1, 99.1 & $6.5 \times 10^{-49}$ & 0.9 & $2.54$ & $2.04$\\
    \bf 1-3 & 147.5, 158.5, 164.8 & 161.8 & 0.73, 0.89 0.96 & 105.1, 99.1 & $1.3 \times 10^{-48}$ & 0.1 & $6.29$ & $6.67 $\\
    \bf 1-4 & 152.7, 158.5, 164.8 & 161.8 & 0.91, 0.89 0.88 & 105.1, 99.1 & $2.26 \times 10^{-48}$ & 0.028 & $7.77$ & $9.75 $\\
    \hline
    \bf 2-1 & 143.2, 173.8, 180.2 & 161.8 & 0.21, 0.68, 0.98 & 81, 76.6  & $2.2 \times 10^{-49}$ & 13.8 & $0.32$ & $0 $\\
    \bf 2-2 & 151.7, 173.8, 180.2 & 177.2 & 0.36, 0.83, 0.98 & 81, 76.6  & $6.2 \times 10^{-49}$ & 2.96 & $1.17$ & $0.51 $\\
    \bf 2-3 & 156.1, 173.8, 180.3 & 177.2 & 0.48, 0.87, 0.98 & 81, 76.6  & $9.9 \times 10^{-49}$ & 0.55 & $2.22$ & $1.83 $\\
    \bf 2-4 & 161.9, 173.8, 180.3 & 177.2 & 0.7, 0.89, 0.96 & 81, 76.6  & $1.92 \times 10^{-48}$ & 0.11 & $4.90$ & $5.19 $\\
    \bf 2-5 & 168.9, 173.8, 180.6& 177.2 & 0.94, 0.89, 0.89 & 81, 76.6  & $3.99 \times 10^{-48}$ & 0.0244 & $6.45$ & $8.74 $\\
    \hline
    \bf 3-1 & 153.7, 184.0, 190.4 & 187.5 & 0.22, 0.72, 0.98 & 67.9, 65  & $3.95 \times 10^{-49}$ & 11.6 & $0.33$ & $0.01 $\\ 
    \bf 3-2 & 160.5, 184.0, 190.4 & 187.5 & 0.33, 0.83, 0.98 & 67.9, 65  & $7.3 \times 10^{-49}$ & 3.45 & $0.91$ & $0.35 $\\ 
     \bf 3-3 & 174.5, 184.0, 190.6 & 187.5 & 0.78, 0.9, 0.96 & 67.9, 65  & $2.76 \times 10^{-48}$ & 0.07 & $5.25$ & $6.36$\\ 
    \bf 3-4 & 180.6, 184.0, 190.9 & 187.5 & 0.97, 0.89, 0.70 & 67.9, 65  & $5.1 \times 10^{-48}$ & 0.018 & $3.72$ & $5.6 $\\ 
    \hline
    \bf 4-1 & 152.3, 194.2, 200.55 & 197.7 & 0.12, 0.46, 0.98 & 57.9, 55.2  & $2.25 \times 10^{-49}$ & 53.3 & $0.08$ & $0 $\\ 
    \bf 4-2 & 162.2, 194.2, 200.6 & 197.7 & 0.21, 0.71, 0.98 & 57.9, 55.2  & $4.7 \times 10^{-49}$ & 13.5 & $0.25$ & $0 $\\ 
    \bf 4-3 & 176.6, 194.2, 200.7 & 197.7 & 0.51, 0.87, 0.98 & 57.9, 55.2  & $1.5 \times 10^{-48}$ & 0.36 & $2.01$ & $1.65 $\\ 
    \bf 4-4 & 185.1, 194.2, 200.7 & 197.7 & 0.82, 0.90, 0.96 & 57.9, 55.2  & $3.4 \times 10^{-48}$ & 0.058 & $5.08$ & $6.20 $\\   
    \bf 4-5 & 190.6, 194.2, 200.7 & 197.7 & 0.97, 0.90, 0.71 & 57.9, 55.2  & $5.6 \times 10^{-48}$ & 0.02 & $3.34$ & $4.99 $\\ 
    \hline
    \bf 5-1 & 169.1, 204.4, 210.8 & 208 & 0.17, 0.66, 0.98 & 49.93, 47.55  & $4.7 \times 10^{-49}$ & 18.8 & $0.14$ & $0 $\\ 
    \bf 5-2 & 179.8, 204.4, 210.8 & 208 & 0.33, 0.84, 0.98 & 49.93, 47.55  & $9.7 \times 10^{-49}$ & 2.7 & $0.70$ & $0.26 $\\
    \bf 5-3 & 184.3, 204.4, 210.8 & 208 & 0.44, 0.88, 0.98 & 49.93, 47.55  & $1.4 \times 10^{-48}$ & 0.56 & $1.33$ & $0.87 $\\
    \bf 5-4 & 194.7, 204.4, 210.8 & 208 & 0.81, 0.90, 0.96 & 49.93, 47.55  & $3.5 \times 10^{-48}$ & 0.0635 & $4.39$ & $5.51 $\\ 
    \bf 5-5 & 200.5, 204.4, 211.3 & 208 & 0.97, 0.90, 0.73 & 49.93, 47.55  & $5.96 \times 10^{-48}$ & 0.02 & $3.11$ & $4.79 $\\
    \hline
    \bf 6-1 & 179.1, 214.5, 221.0 & 218.2 & 0.18, 0.68, 0.98 & 42.6, 40.81  & $5.5 \times 10^{-49}$ & 16.7 & $0.14$ & $0 $\\ 
    \bf 6-2 & 188.6, 214.5, 221.0 & 218.2 & 0.31, 0.84, 0.98 & 42.6, 40.81  & $1.0 \times 10^{-48}$ & 2.85 & $0.56$ & $0.15 $\\     
    \bf 6-3 & 193.3, 214.5, 221.0 & 218.2 & 0.42, 0.88, 0.98 & 42.6, 40.81  & $1.4 \times 10^{-48}$ & 0.62 & $1.04$ & $0.56 $\\
    \bf 6-4 & 201.3, 214.5, 221.0 & 218.2 & 0.68, 0.90, 0.98 & 42.6, 40.81  & $2.7 \times 10^{-48}$ & 0.12 & $2.91$ & $3.47 $\\ 
    \bf 6-5 & 205.9, 214.5, 221.0 & 218.2 & 0.85, 0.91, 0.92 & 42.6, 40.81  & $4.1 \times 10^{-48}$ & 0.052 & $4.06$ & $5.27 $\\ 
    \bf 6-6 & 210.5, 214.5, 221.0 & 218.2 & 0.96, 0.92, 0.74 & 42.6, 40.81  & $6.3 \times 10^{-48}$ & 0.023 & $2.80$ & $4.46 $\\
    \hline
\hline 
  \bottomrule
\end{tabular}
\end{adjustbox}
\endgroup
\caption{Benchmark scenarios, selected from the parameter space in Table~\ref{scanranges}, consistent with all theoretical and LHC constraints and used in our ML-based analysis. Shown are the relevant collider quantities: $m_{\tilde\chi_{1,2,3}^0}$, $\mcharone$, key branching ratios ($\ntrltwo \to \ntrlone \gamma$, $\ntrlthree \to \ntrltwo \gamma$, $\charonepm \to \ntrlone W^{\pm*}$), LO production cross-sections ($pp \to \charonepm \ntrltwo j$, $\charonepm \ntrlthree j$) with $p_T(j) > 100$~GeV and $|\eta_j|<2.5$, as well as DMDD-SI rates, relic density, and ML-based signal significances $Z_{BL}$ and $Z_{MLL}$. Other points are shown in Table~\ref{table:bps3}.}
\label{table:bps2}
\end{table}

\begin{table}[htbp]
\centering
\begingroup
\scriptsize
\setlength{\tabcolsep}{5pt}
\renewcommand{\arraystretch}{1.04}
\setlength{\extrarowheight}{-0.7pt}
\setlength{\aboverulesep}{0.25ex}
\setlength{\belowrulesep}{0.25ex}
\setlength{\abovetopsep}{0.3ex}
\setlength{\belowbottomsep}{0.3ex}
\renewcommand\cellgape{\relax}

\begin{adjustbox}{max width=\textwidth}
\begin{tabular}{@{}c|c c c c c c c c@{}} 
  \toprule
  \textbf{BP} &
  \makecell{$m_{\tilde\chi_1^0,\tilde\chi_2^0,\tilde\chi_3^0}$\\[-2pt]{[GeV]}} &
  \makecell{$m_{\tilde\chi_1^{\pm}}$\\[-2pt]{[GeV]}} &
  \makecell{BR$(\tilde\chi_2^0\!\to\!\tilde\chi_1^0\gamma)$,\\[-2pt]
           BR$(\tilde\chi_3^0\!\to\!\tilde\chi_2^0\gamma)$,\\[-2pt]
           BR$(\tilde\chi_1^{\pm}\!\to\!\tilde\chi_1^0 W^{\pm*})$} &
  \makecell{$\sigma(pp\!\to\!\tilde\chi_1^\pm \tilde\chi_2^0 j)$,\\[-2pt]
           $\sigma(pp\!\to\!\tilde\chi_1^\pm \tilde\chi_3^0 j)$\\[-2pt]
           $(p_T(j)\!>\!100\,\mathrm{GeV},~|\eta_j|\!<\!2.5)$\\[-2pt][fb]} &
  $\sigma_\text{DD}^\text{SI}$ [cm$^{2}$] &
  $\Omega_{\tilde\chi_1^0} h^2$ & $Z_\text{BL}$ & $Z_\text{MLL}$ \\
  \midrule
  \addlinespace
      \bf 7-1 & 197.5, 224.7, 231.2 & 228.4 & 0.3, 0.84, 0.98 & 37.2, 35.6  & $1.0 \times 10^{-48}$ & 3.1 & $0.45$ & $0.08 $\\
    \bf 7-2 & 205.8, 224.7, 231.3 & 228.4 & 0.5, 0.9, 0.98 & 37.2, 35.6  & $1.8 \times 10^{-48}$ & 0.34 & $1.37$ & $1.13 $\\
    \bf 7-3 & 210.7, 224.7, 231.3 & 228.4 & 0.66, 0.91, 0.96 & 37.2, 35.6  & $2.7 \times 10^{-48}$ & 0.134 & $2.49$ & $2.76 $\\
    \bf 7-4 & 215.7, 224.7, 231.5 & 228.4 & 0.84, 0.91, 0.93 & 37.2, 35.6  & $4.1 \times 10^{-48}$ & 0.057 & $3.57$ & $4.71 $\\   
    \bf 7-5 & 218.9, 224.7, 231.5 & 228.4 & 0.93, 0.91, 0.86 & 37.2, 35.6  & $5.6 \times 10^{-48}$ & 0.033 & $3.35$ & $5.05 $\\ 
    \hline
  \bf 8-1 & 213.3, 234.8, 241.5 & 238.6 & 0.43, 0.90, 0.98 & 32.4, 30.9  & $1.6 \times 10^{-48}$ & 0.47 & $0.93$ & $0.61 $\\ 
    \bf 8-2 & 218.5, 234.8, 241.6 & 238.6 & 0.59, 0.91, 0.96 & 32.4, 30.9  & $2.3 \times 10^{-48}$ & 0.19 & $1.73$ & $1.88 $\\
    \bf 8-3 & 223.7, 234.8, 241.7 & 238.6 & 0.77, 0.91, 0.94 & 32.4, 30.9  & $3.6 \times 10^{-48}$ & 0.081 & $2.85$ & $3.67$\\ 
    \bf 8-4 & 228.8, 234.8, 241.8 & 238.6 & 0.93, 0.91, 0.87 & 32.4, 30.9  & $5.7 \times 10^{-48}$ & 0.081 & $3.09$ & $4.63 $\\
    \hline
    \bf 9-1 & 222.5, 245.0, 251.7 & 248.9 & 0.41, 0.90, 0.98 & 28.4, 27.1  & $1.6 \times 10^{-48}$ & 0.51 & $0.75$ & $0.48$\\ 
    \bf 9-2 & 227.9, 245.0, 251.7 & 248.9 & 0.57, 0.91, 0.96 & 28.4, 27.1  & $2.3 \times 10^{-48}$ & 0.21 & $1.47$ & $1.44$\\ 
    \bf 9-3 & 235.1, 245.0, 251.7 & 248.9 & 0.82, 0.91, 0.92 & 28.4, 27.1  & $4.2 \times 10^{-48}$ & 0.07 & $2.84$ & $3.80$\\ 
    \bf 9-4 & 240.3, 245.0, 252.1 & 248.9 & 0.96, 0.91, 0.76 & 28.4, 27.1  & $6.8 \times 10^{-48}$ & 0.03 & $2.10$ & $3.51$\\   
    \hline
    \bf 10-1 & 235.4, 255.1, 261.9 & 259.1 & 0.5, 0.90, 0.98 & 25, 23.9  & $2.02 \times 10^{-48}$ & 0.3 & $0.99$ & $0.66$\\ 
    \bf 10-2 & 239.2, 255.1, 262.0 & 259.1 & 0.62, 0.91, 0.96 & 25, 23.9  & $2.64 \times 10^{-48}$ & 0.17 & $1.57$ & $1.59$\\
    \bf 10-3 & 244.8, 255.1, 262.0 & 259.1 & 0.81, 0.91, 0.93 & 25, 23.9  & $4.2 \times 10^{-48}$ & 0.07 & $2.55$ & $3.42$\\ 
    \bf 10-4 & 250.3, 255.1, 262.3 & 259.1 & 0.96, 0.91, 0.75 & 25, 23.9  & $6.7 \times 10^{-48}$ & 0.03 & $1.88$ & $3.01$\\
    \hline
    \bf 11-1 & 244.8, 265.2, 272.1 & 269.3 & 0.48, 0.9, 0.97 & 22, 21.1  & $2.0 \times 10^{-48}$ & 0.32 & $0.82$ & $0.58$\\ 
    \bf 11-2 & 250.6, 265.2, 272.1 & 269.3 & 0.67, 0.91, 0.95 & 22, 21.1  & $3.0 \times 10^{-48}$ & 0.13 & $1.66$ & $1.82$\\
    \bf 11-3 & 254.5, 265.2, 272.2 & 269.3 & 0.8, 0.91, 0.93 & 22, 21.1  & $4.1 \times 10^{-48}$ & 0.077 & $2.14$ & $2.85$\\
    \bf 11-4 & 258.3, 265.2, 272.4 & 269.3 & 0.91, 0.91, 0.87 & 22, 21.1  & $5.8 \times 10^{-48}$ & 0.045 & $2.24$ & $3.47$\\
    \bf 11-5 & 262.1, 265.2, 272.6 & 269.3 & 0.98, 0.90, 0.62 & 22, 21.1  & $8.1 \times 10^{-48}$ & 0.026 & $1.12$ & $1.69$\\ 
    \hline
     \bf 12-1 & 252.1, 275.3, 282.3 & 279.5 & 0.41, 0.9, 0.98 & 19.5, 18.7  & $1.7 \times 10^{-48}$ & 0.47 & $0.57$ & $0.27$\\  
     \bf 12-2 & 256.1, 275.3, 282.3 & 279.5 & 0.52, 0.91, 0.96 & 19.5, 18.7  & $2.2 \times 10^{-48}$ & 0.26 & $0.94$ & $0.74$\\
     \bf 12-3 & 260.2, 275.3, 282.3 & 279.5 & 0.65, 0.91, 0.95 & 19.5, 18.7  & $3.0 \times 10^{-48}$ & 0.14 & $1.42$ & $1.54$\\ 
     \bf 12-4 & 264.2, 275.3, 282.4 & 279.5 & 0.79, 0.91, 0.93 & 19.5, 18.7  & $4.1 \times 10^{-48}$ & 0.083 & $2.00$ & $2.59$\\ 
     \bf 12-5 & 268.2, 275.3, 282.5 & 279.5 & 0.91, 0.91, 0.84 & 19.5, 18.7  & $5.8 \times 10^{-48}$ & 0.048 & $1.96$ & $3.06$\\
     \hline
     \bf 13-1 & 265.5, 285.4, 292.5 & 289.7 & 0.51, 0.91, 0.96 & 17.2, 16.6  & $2.2 \times 10^{-48}$ & 0.27 & $0.80$ & $0.51$\\ 
     \bf 13-2 & 271.8, 285.4, 292.6 & 289.7 & 0.71, 0.91, 0.95 & 17.2, 16.6  & $3.4 \times 10^{-48}$ & 0.117 & $1.55$ & $1.80$\\ 
     \bf 13-3 & 276.0, 285.4, 292.6 & 289.7 & 0.85, 0.91, 0.91 & 17.2, 16.6  & $4.8 \times 10^{-48}$ & 0.067 & $1.90$ & $2.65$\\ 
     \bf 13-4 & 280.1, 285.4, 292.8 & 289.7 & 0.95, 0.91, 0.81 & 17.2, 16.6  & $7 \times 10^{-48}$ & 0.04 & $1.52$ & $2.43$\\ 
         \hline
     \bf 14-1 & 277.1, 295.4, 302.7 & 299.9 & 0.56, 0.91, 0.96 & 15.4, 14.2  & $2.4 \times 10^{-48}$ & 0.22 & $0.58$ & $0.21$\\ 
     \bf 14-2 & 285.7, 295.4, 302.7 & 299.9 & 0.85, 0.91, 0.90 & 15.4, 14.2  & $4.8 \times 10^{-48}$ & 0.072 & $0.84$ & $0.53$\\ 
     \bf 14-3 & 290, 295.4, 302.7 & 299.9 & 0.95, 0.91, 0.75 & 15.4, 14.2  & $7 \times 10^{-48}$ & 0.042 & $0.76$ & $0.43$\\ 
              \hline
     \bf 15-1 & 289.3, 305.5, 313.2 & 310 & 0.63, 0.91, 0.94 & 13.7, 13.2  & $2.9 \times 10^{-48}$ & 0.165 & $0.60$ & $0.24$\\ 
     \bf 15-2 & 295.5, 305.5, 313.2 & 310 & 0.84, 0.91, 0.90 & 13.7, 13.2  & $4.8 \times 10^{-48}$ & 0.076 & $0.73$ & $0.35$\\ 
     \bf 15-3 & 300, 305.5, 313.2 & 310 & 0.95, 0.91, 0.75 & 13.7, 13.2  & $6.9 \times 10^{-48}$ & 0.044 & $0.69$ & $0.43$\\ 
                   \hline
     \bf 16-1 & 300, 315.5, 323.1 & 320 & 0.66, 0.91, 0.93 & 12.2, 11.75  & $3 \times 10^{-48}$ & 0.151 & $0.57$ & $0.26$\\ 
     \bf 16-2 & 300, 315.5, 323.1 & 320 & 0.86, 0.91, 0.86 & 12.2, 11.75  & $5 \times 10^{-48}$ & 0.071 & $0.66$ & $0.31$\\ 
     \bf 16-3 & 310.7, 315.5, 323.1 & 320 & 0.96, 0.91, 0.62 & 12.2, 11.75  & $7.5 \times 10^{-48}$ & 0.042 & $0.53$ & $0.21$\\ 
                        \hline
     \bf 17-1 & 309.7, 325.5, 333.2 & 330.4 & 0.66, 0.91, 0.93 & 10.6, 10.95  & $3 \times 10^{-48}$ & 0.155 & $0.50$ & $0.17$\\ 
     \bf 17-2 & 313.6, 325.5, 333.2 & 330.4 & 0.78, 0.91, 0.90 & 10.6, 10.95  & $4 \times 10^{-48}$ & 0.10 & $0.57$ & $0.27$\\ 
     \bf 17-3 & 320.6, 325.5, 333.2 & 330.4 & 0.96, 0.91, 0.72 & 10.6, 10.95  & $3 \times 10^{-48}$ & 0.045 & $0.57$ & $0.21$\\      
\hline 
  \bottomrule
\end{tabular}
\end{adjustbox}
\endgroup
\caption{Continuation of Table~\ref{table:bps2}.}
\label{table:bps3}
\end{table}

\newpage

\bibliographystyle{JHEP} 
\bibliography{biblio.bib}

\end{document}